\UseRawInputEncoding 
\documentclass[preprint,12pt,fleqn]{elsarticle}
\usepackage{amsmath}
\usepackage{amsthm}
\usepackage{amssymb}
\usepackage{eucal}
\usepackage{mathrsfs}
\usepackage{subfigure}
\usepackage{mathtools} %add
\usepackage{nccmath} %add
\usepackage{fix-cm}
\usepackage{bm}

\def \div{\mbox{\rm div}}

\def \be {\beta}
\def \ga {\gamma}
\def \Ga {\Gamma}

\def \ep {\varepsilon}

\def \Om {\Omega}

\begin{document}

\begin{frontmatter}

\title{Topology optimization considering the distortion in additive manufacturing}
\author[mymainaddress]{Takao Miki}
\author[mysecondaryaddress]{Takayuki Yamada\corref{mycorrespondingauthor}}
\cortext[mycorrespondingauthor]{Corresponding author}
\ead{t.yamada@mech.t.u-tokyo.ac.jp}
\address[mymainaddress]{Osaka Research Institute of Industrial Science and Technology, 7-1, Ayumino-2, Izumi-city, Osaka, 594-1157, Japan}
\address[mysecondaryaddress]{Department of Strategic Studies, Institute of Engineering Innovation, The University of Tokyo, 11-16, Yayoi-2, Bunkyo-ku, Tokyo, 113-8656, Japan}

\begin{abstract}
	Additive manufacturing is a free-form manufacturing technique in which parts are built in a layer-by-layer manner.
	Laser powder bed fusion is one of the popular techniques used to fabricate metal parts.
	However, it induces residual stress and distortion during fabrication that adversely affects the mechanical properties and dimensional accuracy of the manufactured parts.
	Therefore, predicting and avoiding the residual stress and distortion are critical issues.
	In this study, we propose a topology optimization method that accounts for the distortion.
	First, we propose a computationally inexpensive analytical model for additive manufacturing that uses laser powder bed fusion and formulated an optimization problem.
	Next, we approximate the topological derivative of the objective function using an adjoint variable method that is then utilized to update the level set function via a time evolutionary reaction-diffusion equation. Finally, the validity and effectiveness of the proposed optimization method was established using two-dimensional design examples.
\end{abstract}
\end{frontmatter}
\section{Introduction}
Topology optimization \cite{bendsoe1988generating,bendsoe1989optimal} is employed in the design of high-performance structures.
However, these structures often have complicated shapes, and cannot be manufactured directly by conventional technologies such as machining and molding.
To solve this problem, manufacturability needs to be considered in the optimization procedure.
For instance, topology optimization methods such as controlling the length scale \cite{allaire2016thickness,yamada2018thickness} or avoiding internal voids and undercut shapes in molding \cite{xia2010level,allaire2016molding,sato2017manufacturability} have been proposed to realize shapes optimized for manufacturing.
Metal additive manufacturing (AM) is attracting attention as a free-form manufacturing technique that can build complicated shapes in a bottom-up, layer-by-layer manner.
In particular, laser powder-bed fusion (LPBF) is utilized in various applications including aerospace, automotive, medical, energy, etc.
However, manufacturability issues such as overhang constraints need to be considered in LPBF, and several approaches have been proposed to incorporate it into topology optimization \cite{leary2014optimal,langelaar2016topology,gaynor2016topology,allaire2017structural,wang2018level}.
In addition, residual stress and distortion must be considered to ensure successful fabrication.
In this study, we focus on topology optimization that reduces the distortion induced by AM during the optimization procedure.

In the LPBF process, materials are melted and solidified by rapid local heating and cooling.
This heating and cooling cycle generates thermal, plastic, and transformation strains, which cause residual stress and distortion.
These adversely affect the strength and dimensional accuracy of the manufactured parts.
Therefore, predicting and avoiding residual stress and distortion are crucial issues.

Residual stress and distortion in AM have been studied extensively using experimental and numerical methods \cite{kruth2004selective,mercelis2006residual,van2013investigation,wu2014experimental,fergani2017analytical,li2018residual,bugatti2018limitations}.
Existing numerical methods can be summarized into two approaches: the thermal-elastic-plastic analysis method \cite{ueda1971analysis} and the inherent strain method \cite{ueda1975new,murakawa1996prediction} that was originally used in welding.
The thermal-elastic-plastic analysis method is a coupled analysis that combines heat transfer analysis and elastic or elastic-plastic analysis.
Heat transfer analysis uses a detailed model with a moving heat source and the micro-scale layer thickness makes it computationally expensive.
In the elastic-plastic analysis, it is necessary to consider the nonlinearity of the material properties and the solid-liquid phase transition.
To address this, many simplified methods have been proposed to predict the residual stress and distortion on a part-scale \cite{papadakis2014numerical,hodge2014implementation,li2016multiscale,mukherjee2017improved,denlinger2017thermomechanical,chiumenti2017numerical,li2017efficient}.
However, it is difficult to replace the moving heat source with a simplified heat source, and the effectiveness of these analytical models has not been firmly established.
In contrast, with the inherent strain method, the strain component is identified experimentally from the fabricated part and applied to an analysis domain.
Then, the residual stress and distortion are obtained by linear elastic analysis.
This makes it is computationally inexpensive compared to the thermal-elastic-plastic analysis method in that it does not require coupled and nonlinear analyses.
The effectiveness and validity of the inherent strain method have been established previously by experimental verification \cite{keller2014new,setien2019empirical,chen2019inherent,liang2019modified,prabhune2020fast}.
Based on these numerical analysis methods, topology optimization methods have been proposed that reduce the residual stress and distortion in AM.
Wildman et al. \cite{wildman2017topology} proposed a multi-objective topology optimization method using a solid isotropic material with penalization scheme and defined the objective function for mean compliance and reducing the induced distortion.
A thermo-elastic element-birth model is thus proposed in which the elements are sequentially activated to simulate the moving heat source.
Allaire et al. \cite{allaire2018taking} proposed a layer-by-layer thermo-elastic analysis model and incorporated it into a level set-based topology optimization to minimize the distortion and residual stress in each layer with an objective function.
In this approach, the optimal configuration obtained from a mean compliance minimization problem is reoptimized as the initial shape.
By introducing the mean compliance and a predefined stress threshold as constraints to the optimization problem, the optimization method succeeded in minimizing the mean compliance and reducing the residual stress.
However, both approaches are computationally intensive, owing to the use of coupled analysis in the AM analytical model.
This makes the methods inefficient for topology optimization that requires iterative procedures.
In addition, the validity of the proposed analytical models have not been established.

In this study, we present a topology optimization method by considering the part distortion in AM, using a computationally inexpensive analytical model.
Specifically, we propose an analytical model based on the inherent strain method that realizes computationally inexpensive and accurate numerical analysis to predict the residual stress and distortion.
The proposed model is incorporated into a topology optimization process to obtain optimal high-performance configurations that account for the part distortion in AM.

The remainder of this paper is organized as follows: In section \ref{sec:2}, we propose an analytical model based on the inherent strain method for predicting the residual stress and distortion in the AM building process.
Section \ref{sec:3} explains an identification method of the inherent strain component. 
Section \ref{sec:4} presents an experimental validation of the proposed analytical model.
Section \ref{sec:5} describes the level set-based topology optimization method, in which the level set function is updated using the reaction-diffusion equation.
In Section \ref{sec:6}, we formulate an optimization problem that incorporates the reduction of part distortion in the topology optimization procedure, and derive an approximate topological derivative using variational analysis and the adjoint variable method.
In Section \ref{sec:7}, we constructed an optimization algorithm for the topology optimization using the finite element method (FEM).
Section \ref{sec:8} presents two-dimensional design examples to demonstrate the validity of the proposed optimization method. 
Lastly, Section \ref{sec:9} brings the study to a close with conclusions.
All numerical calculations were executed in FreeFEM++\cite{MR3043640}.
\section{Analytical model for AM}\label{sec:2}
In this section, the layer-by-layer type building process in AM is modeled using the inherent strain method, a type of linear elastic analysis.
The reason for using such a method is that optimization requires iterative calculations, therefore, computationally inexpensive models are desirable.
To simplify the model, we assume that a constant strain is associated with each layer.
\subsection{Inherent strain method}
In the AM building process, inelastic strains such as thermal, plastic, and phase transformation strains are induced in the parts through the melting-solidification cycle of the material.
The sum of these inelastic strains is called the inherent strain \cite{ueda1975new,murakawa1996prediction}.
The total strain $\bm{\ep}(\bm{u})$ is defined by:  $\bm{\ep}(\bm{u})\coloneqq\frac{1}{2}(\nabla\bm{u}+(\nabla\bm{u})^{\top})$, and it can be divided into the elastic strain $\bm{\ep}^{el}$ and inherent strain $\bm{\ep}^{inh}$.
The inherent strain $\bm{\ep}^{inh}$ and the Cauchy stress tensor $\bm{\sigma}$ are defined as:
\begin{align}
&\bm{\ep}(\bm{u})=\bm{\ep}^{el}+\bm{\ep}^{inh},\label{eq:am1}\\
&\bm{\sigma}=\mathbb{C}\bm{\ep}^{el}=\mathbb{C}\bm{\ep}(\bm{u})-\mathbb{C}\bm{\ep}^{inh},\label{eq:am2}
\end{align}
where $\mathbb{C}$ is the fourth-order elasticity tensor.
For an isotropic elastic material, the above tensor is given by:
\begin{equation}
\mathbb{C}_{i j k l}=E\left(\frac{\nu}{(1+\nu)(1-2 \nu)} \delta_{i j} \delta_{k l}+\frac{1}{2(1+\nu)}\left(\delta_{i k} \delta_{j l}+\delta_{i l} \delta_{j k}\right)\right),
\label{eq:am3}
\end{equation}
with Young's modulus $E$, the Poisson's ratio $\nu$ and the Kronecker delta $\delta_{i j}$.
The residual stress and distortion can be predicted using a linear analysis by applying the inherent strain in a layer-by-layer manner.
\begin{figure}[htbp]
\begin{center}
	\includegraphics[width=13.0cm]{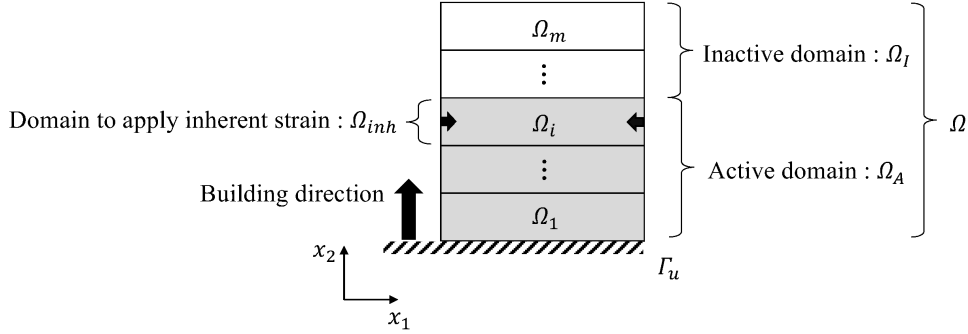}
	\caption{Domains and boundary in the middle of the building process.}
	\label{fig:analytical model}
\end{center}
\end{figure}
As shown in Fig. \ref{fig:analytical model}, we consider an analysis domain $\mathit{\Om}$, divided into $m$ layers with a fixed thickness in the building direction.
The analysis domain $\mathit{\Om}$ is defined by each domain $\mathit{\Om}_{i}$ for $1\leq i\leq m$ as follows:
\begin{equation}
\mathit{\Om} = \mathit{\Om_{1}} \cup{\ldots} \cup{\mathit{\Om_{i}}} \cup{\ldots} \cup{\mathit{\Om_{m}}}.
\label{eq:am4}
\end{equation}
Here, we introduce three subdomains as: the active domain $\mathit{\Om_{A}}$, the inactive domain $\mathit{\Om_{I}}$, and the domain $\mathit{\Om_{inh}}$ to which the inherent strain is applied.
The subdomain region depends on the domain number $i$, and each subdomain is defined as:
\begin{align}
&\mathit{\Om_{A}} = \mathit{\Om_{1}} \cup{\ldots} \cup{\mathit{\Om_{i}}}, \label{eq:am5}\\
&\mathit{\Om_{I}} = \mathit{\Om} \setminus \mathit{\Om_{A}}, \label{eq:am6}\\
&\mathit{\Om_{inh}} = \mathit{\Om_{i}} \subset{\mathit{\Om_{A}}}. \label{eq:am7}
\end{align}
The active domain $\mathit{\Om_{A}}$ is filled with an elastic material, and a fixed displacement boundary condition is applied to the bottom of the domain $\mathit{\Ga_{u}}$.
The mechanical unknown of this model is the displacement field.
The displacement $\bm{u}_{i} \in \mathcal{U} $ with the inherent strain applied to the domain $\mathit{\Om_{inh}}$ is governed by the equations of linear elasticity as follows:
\begin{align}
\begin{cases}
\hspace{2mm}-\div(\bm{\sigma}_{i})=0\hspace{5mm}&\text{in}\hspace{2mm}\mathit{\Om_{A}},\\
\hspace{2mm}\bm{\sigma}_{i}=\mathbb{C}\bm{\ep}(\bm{u}_{i})-\mathbb{C}\bm{\ep}^{inh},\hspace{5mm}&\\
\hspace{2mm}\bm{u}_{i}=\bm{0}\hspace{5mm}&\text{on}\hspace{2mm}\mathit{\Ga_{u}},\\
\hspace{2mm}-\bm{\sigma}_{i}\cdot\bm{n}=\bm{0}\hspace{5mm}&\text{on}\hspace{2mm}\partial \mathit{\Om_{A}} \setminus \mathit{\Ga_{u}},
\label{eq:am8}
\end{cases}
\end{align}
\begin{equation}
\mathcal{U}:=\left\{\bm{u}_{i} \in H_{\Ga_{u}}^{1}(\mathit{\Om}_{A})^{N}, \hspace{1mm} \bm{u}_{i}=\bm{0} \hspace{1mm} \text { on } \hspace{1mm} \mathit\Ga_{u}\right\},
\label{eq:am9}
\end{equation}
for all indices $i = 1,2,\ldots,m$, where $N$ is the number of spatial dimensions.
The inherent strain $\bm{\ep}^{inh}$ at the domain $\mathit{\Om_{inh}}$ is defined as:
\begin{equation}
\bm{\ep}^{inh} ( \bm{x} ) = \left\{ \begin{array} { l l } { \bm{\ep}^{inh} } & { \text { for } \bm{x} \in \mathit{\Omega_{inh}} }, \\ { \bm{0} } & { \text { otherwise } }, \end{array} \right.
\label{eq:am10}
\end{equation}
where $\bm{x}$ represents a point located in $\mathit{\Omega_{A}}$.
The method for identifying the inherent strain component of Eq. \ref{eq:am10} is described in the next section.
\subsection{AM building process model}
The AM building process is represented by various activation strategies in finite element modeling \cite{chiumenti2017numerical,setien2019empirical,wildman2017topology,allaire2018taking,foteinopoulos2018thermal}.
In this study, we use a method in which the inactive domain is activated sequentially from the bottom domain, and the inherent strain is applied to each activated domain.
To solve the analysis domain $\mathit{\Om}$ using FEM, we use an ersatz material approach in which the inactive domain $\mathit{\Om_{I}}$ is occupied by a structural material with a relatively small Young's modulus.

Our AM building process algorithm is as follows:
\begin{description}
\item[Step1.] Inactivate all domains in the analysis domain divided into $m$ layers.
\item[Step2.] The domains are activated in sequence from the bottom domain, i.e., the activated domain is replaced by the original Young's modulus, and the inherent strain is applied after this.
\item[Step3.] The displacement field $\bm{u}_{i}$ defined in Eqs. \ref{eq:am8} and \ref{eq:am9} is solved using FEM.
\item[Step4.] If all domains are activated, the procedure is terminated; otherwise, return to the second step.
\end{description}
The analysis domain is divided into fewer layers than the actual number of layers during manufacturing, owing to the computational costs. 
This multi-scale modeling enables the prediction of the part-scale residual stress and distortion. 
%The final residual stress, $\bm{\sigma}$, and distortion, $\bm{u}$, of the part are the sums of the residual stress $\bm{\sigma}_{i}$ and distortion $\bm{u}_{i}$, respectively. The residual stress and distortion were solved for each active domain $\mathit{\Om_{A}}$, as follows: \cite{liang2019modified}
The final residual stress $\bm{\sigma}$ and distortion $\bm{u}$ of the part are the sums of the residual stress $\bm{\sigma}_{i}$ and distortion $\bm{u}_{i}$ solved for each active domain $\mathit{\Om_{A}}$, as follows: \cite{liang2019modified}
\begin{align}
\bm{\sigma} &= \sum_{i=1}^{m}\bm{\sigma}_{i} \hspace{5mm}\text { for } \bm{x} \in \mathit{\Omega_{A}}, \label{eq:am11}\\
\bm{u} &= \sum_{i=1}^{m}\bm{u}_{i} \hspace{5mm}\text { for } \bm{x} \in \mathit{\Omega_{A}}. \label{eq:am12}
\end{align}
\section{Inherent strain identification method}\label{sec:3}
\subsection{Experimental procedure}
\begin{figure}[htbp]
\begin{center}
	\includegraphics[width=10.0cm]{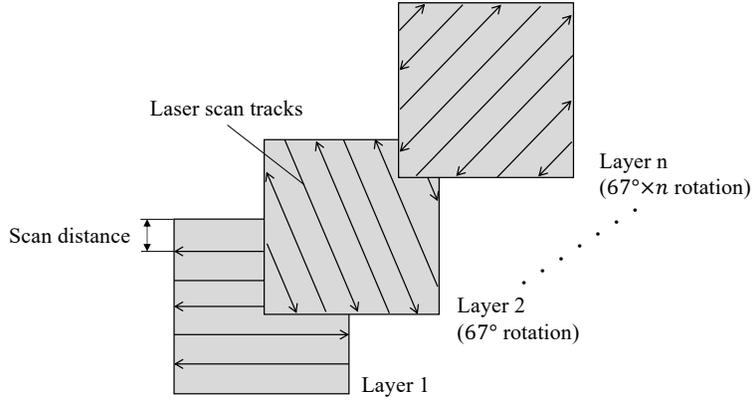}
	\caption{Schematic illustration of the laser scanning pattern.}
	\label{fig:Scanningpattern}
\end{center}
\end{figure}

The unknown inherent strain component can be identified based on the measurements of a deformation caused by releasing the elastic strain \cite{ueda1975new,ueda1979new}.
The elastic strain needs to be released such that the inherent strain is unchanged, using devices such as wire electric discharge machines (WEDM).
%The elastic strain needs to be released by manufacturing technology that does not change the inherent strain, such as wire electric discharge machines (WEDM).
Specimens were fabricated using an LPBF machine (EOSINT M280, EOS GmbH) equipped with a 400 W fiber laser (beam diameter: approximately 0.1 mm).
\begin{figure}[htbp]
\begin{center}
	\centering
	\includegraphics[width=13.0cm]{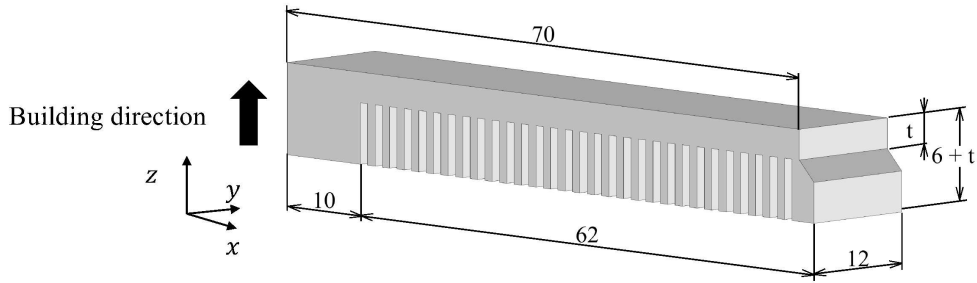}
	\caption{Cantilever specimen with the dimensions in mm.}
	\label{fig:CantileverSpecimen}
\end{center}
\end{figure}
Figure. \ref{fig:CantileverSpecimen} shows the geometry of the specimen and its dimensions.
The beam thickness (3 mm) is represented by $t$.
%$t$ denotes the beam thickness of 3 mm.
The specimens made of AlSi10Mg (EOS Aluminum, EOS GmbH) were fabricated on the substrate with an argon atmosphere.
The laser scanning parameters used an original EOS parameter set adjusted for laser power, scan speed, and scan distance.
The laser scanning pattern was rotated by $67^\circ$ layer-by-layer, as shown in Fig. \ref{fig:Scanningpattern}, and the constant material layer thickness was 0.03 mm.

\begin{figure}[htbp]
\begin{center}
	\includegraphics[width=7.0cm]{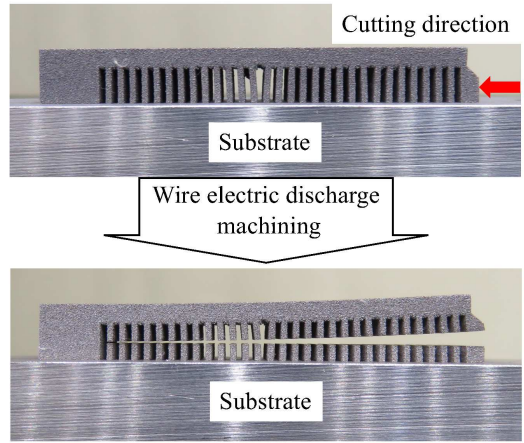}
	\caption{Schematic diagram of the specimens being cut.}
	\label{fig:SpecimenWEDM}
\end{center}
\end{figure}

After fabrication, the specimen was partially cut with WEDM at a height of 3 mm from the substrate, leaving the left column attached to the substrate, as shown in Fig. \ref{fig:SpecimenWEDM}.
The elastic strain is released upon cutting, leading to a large deformation of the specimens.
The vertical deformation was calculated from the difference in the top surfaces before and after cutting, which were measured using an optical three-dimensional scanner (ATOS Core, GOM GmbH).
The inherent strain component was identified from the experimental and numerical results.
\subsection{Finite element modeling}
\begin{figure}[htbp]
\begin{center}
	\centering
	\subfigure[]{\includegraphics[width=13.0cm]{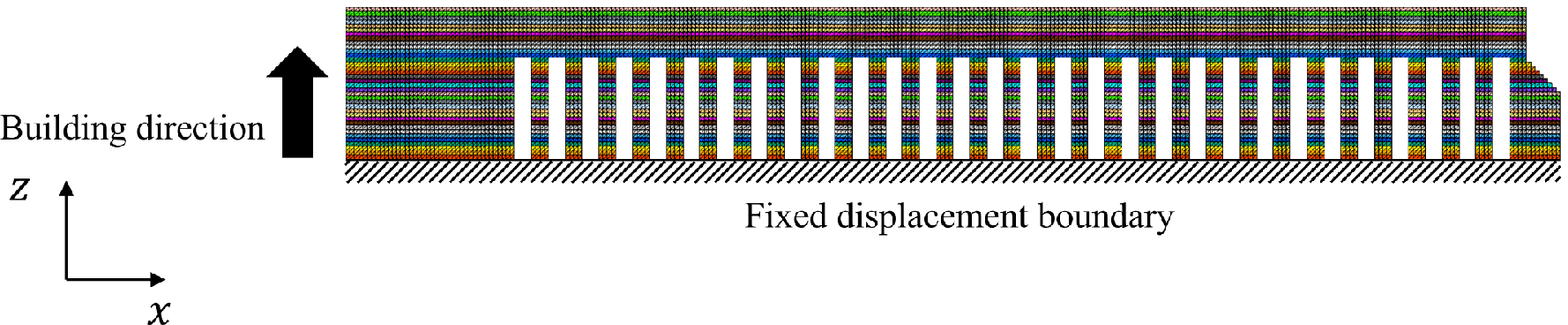}}
	%		\hspace{0.1cm}
	\subfigure[]{\includegraphics[width=13.0cm]{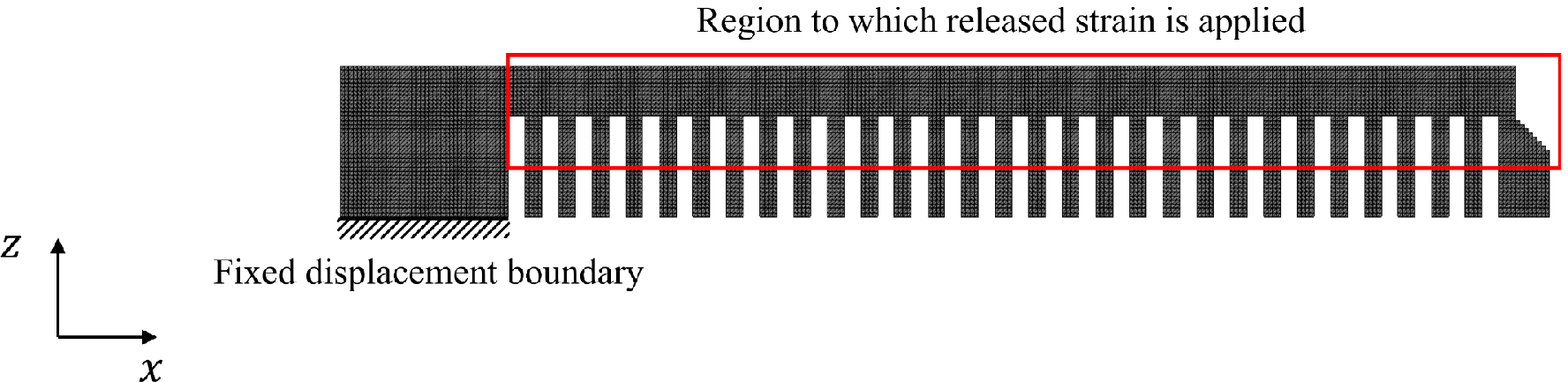}}
	\caption{Boundary conditions and mesh model of the cantilever beam $t = 3$ mm: (a) AM building process model and (b) cutting model.}
	\label{fig:FemModel}
\end{center}
\end{figure}
Figure \ref{fig:FemModel} shows the mesh and analysis conditions for the AM building and cutting processes.
In the AM building process, the displacements of the bottom surfaces are fixed, where the color difference represents each layer to which the inherent strain is applied sequentially.
%In the AM building process, the displacements of the bottom surfaces are fixed.
%The color difference represents each layer to which the inherent strain is applied sequentially.
The elasticity problem is solved layer-by-layer according to the AM building process algorithm to determine the elastic strain of the specimen.
In the cutting process, only the bottom surface of the left column is fixed, and the elastic strain released by cutting is applied to the part to determine the deformation.
The region of the applied elastic strain is demarcated by the red rectangular frame (3 mm height from the bottom and unfixed area).
In this study, the layer thickness was discretized with an element size of 0.25 mm for the part-scale analysis, which is approximately 10 times the actual material layer thickness in the fabrication.
%The mesh applied to discretization is a second-order tetrahedral with 780,861 nodes and 4,138,200 elements.
%The mesh applied to discretization is 2,114,496 second-order tetrahedral elements.
The mesh applied to discretization consists of 2,114,496 second-order tetrahedral elements.
%The mesh model have 780,861 nodes and 4,138,200 elements. 
%The mesh applied to the discretization is a second-order tetrahedral element.
%In addition, the material properties for the active domains are the Young's modulus = 75 GPa and Poisson's ratio = 0.34.
%In addition, the Young's modulus for the active and inactive domains was set to 75 GPa and 0.01 MPa, respectively, and the Poisson's ratio was set 0.34.}
In addition, the Young's modulus for the active and inactive domains were set to 75 GPa and 0.01 MPa, respectively and Poisson's ratio was set to 0.34.
\subsection{Identification procedure}
The inherent strain component has been identified under two assumptions.
First, the fabricated parts shrink isotropically as the laser scan pattern rotates layer-by-layer.
Therefore, the in-plane components are assumed to have the same value \cite{setien2019empirical}.
Second, because the layer thickness is extremely small compared to the part size, the building direction component is assumed to be zero \cite{bugatti2018limitations,setien2019empirical}.
%Second, since the layer thickness is sufficiently small compared to the part size, the building direction component can be assumed to be zero \cite{bugatti2018limitations,setien2019empirical}.
The in-plane inherent strain components were obtained through the minimization of the residual sum of squares of the experimental and numerical vertical deformation results.
\begin{figure}[htbp]
\begin{center}
	\includegraphics[width=10.0cm]{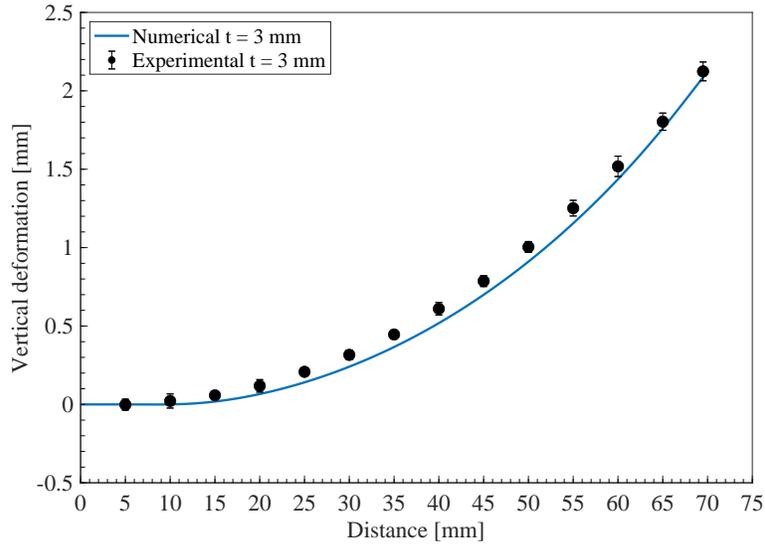}
	\caption{Comparison of the experimental and numerical results of the vertical deformation after the inherent strain identification procedure.}
	\label{fig:IdentificationResult}
\end{center}
\end{figure}
Figure \ref{fig:IdentificationResult} shows the comparison between the numerical and experimental result of the vertical deformation of the top surface after minimization.
The experimental result was plotted for the vertical deformation of the top surface at cross-sections measured every 5 mm along the $x$-direction.
The error bars represent the minimum and maximum values in the cross-section.
The inherent strain component obtained by this procedure is as follows:
%In this case, the in-plane inherent strain components obtained by this procedure is as follows:
\begin{equation}
\bm\ep^{i n h}
=\left\{\begin{array} {l} 
\ep_{x} \\
\ep_{y} \\
\ep_{z}
\end{array}\right\}
=\left\{\begin{array}{c} 
-0.250 \\
-0.250 \\
0
\end{array}\right\}
.\label{eq:am13}
\end{equation}
\section{Validation of analytical model}\label{sec:4}
\begin{figure}[htbp]
\begin{center}
	\centering
	\subfigure[]{\includegraphics[width=10.0cm]{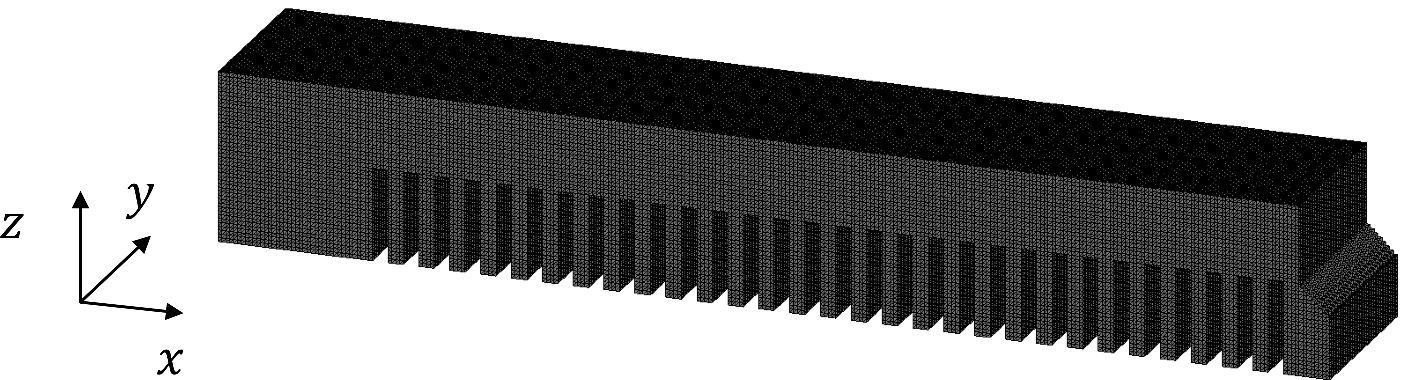}}
	%		\hspace{0.1cm}
	\subfigure[]{\includegraphics[width=10.0cm]{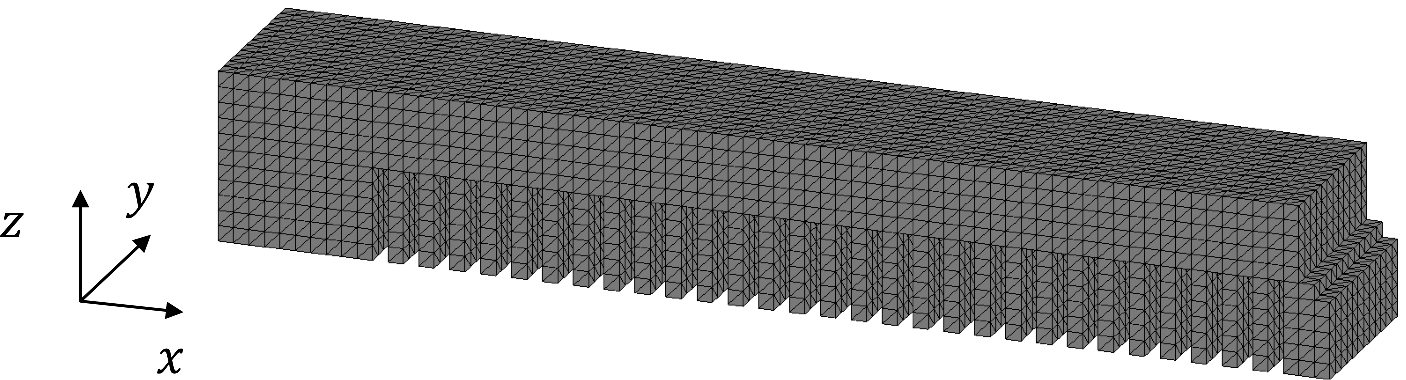}}
	\caption{Mesh model of the cantilever specimen $t = 5$ mm: (a) element size of 0.25 mm; (b) element size of 1.0 mm.}
	\label{fig:FemModelmesh}
\end{center}
\end{figure}

The validity of the identified inherent strain and the proposed analytical model were verified through two types of experiments.
First, to verify the validity of predicting the distortion, we prepared a cantilever specimen with the beam thickness t shown in Fig. \ref{fig:CantileverSpecimen} set to 5 mm, then presents a comparison of the experimental and numerical results of the vertical deformation after cutting the specimen.
%First, in order to verify the validity of predicting the distortion, we prepared a cantilever specimen with the beam thickness $t$ set to 5 mm in Fig. \ref{fig:CantileverSpecimen} presents a comparison of the experimental and numerical results of the vertical deformation after cutting the specimen.
The cantilever specimen made of AlSi10Mg was fabricated under the same manufacturing conditions the beam with thickness $t = 3$ mm.
After fabrication, the specimen was cut by WEDM, and the deformation was measured with the optical three-dimensional scanner.
%In the numerical analysis, we prepared two models: in the one model the element size was discretized at 0.25 mm, while in the other model at 1 mm.
%This was done to compare the accuracy and computational time (see Fig. \ref{fig:FemModelmesh}). 
In the numerical analysis, we prepared two models that the element sizes per layer were discretized at 0.25 mm and 1 mm to compare the accuracy and computational time, as shown in Fig. \ref{fig:FemModelmesh}.
%In the numerical analysis, we prepared two models wherein the element sizes per layer were discretized at 0.25 mm and 1 mm to compare the accuracy and computational time, as shown in Fig. \ref{fig:FemModelmesh}.
%Furthermore, 
The meshes of the two models consist of 2,759,616 and 43,272 second-order tetrahedral elements, respectively.
%	In the numerical analysis, to compare the accuracy and computing time, we prepared models discretized to 0.5 mm and 1 mm with respect to the building direction in Fig.\ref{fig:FemModelmesh}.
The same boundary condition, material properties, and inherent strain as in the beam with thickness $t = 3$ mm were used.
%The numerical analysis used the same boundary conditions, the mesh size, material properties, and the inherent strain as the beam thickness $t=3$ mm.

\begin{figure}[htbp]
\begin{center}
	\includegraphics[width=10.0cm]{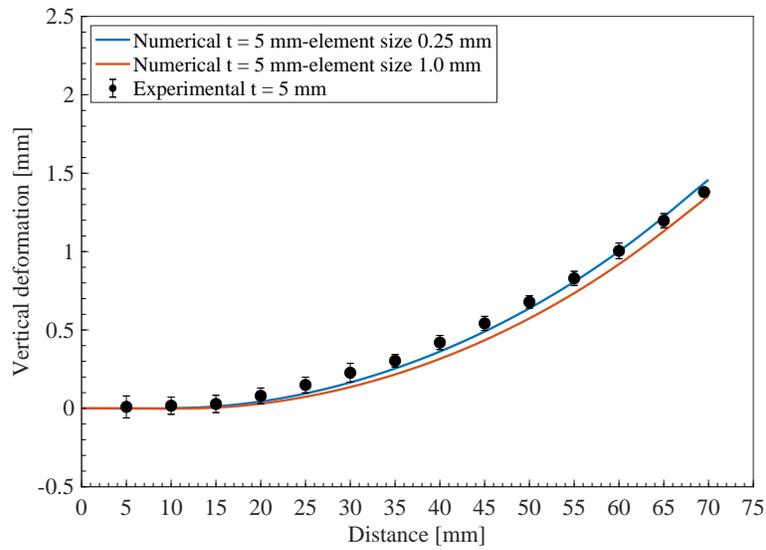}
	\caption{Comparison of the experimental and numerical results of the vertical deformation after cutting with WEDM.}
	\label{fig:t5deformation}
\end{center}
\end{figure}

Figure \ref{fig:t5deformation} shows the comparison between the numerical and experimental results for the vertical deformation of the top surface.
The experimental result was plotted for the vertical deformation of the top surface at cross-sections measured every 5 mm along the $x$-direction.
The error bars represent the minimum and maximum values in the cross-section.
The numerical results for the two different element sizes are consistent with the experimental measurements.
This verification demonstrates that the part-scale distortion induced in the AM building process can be accurately predicted for both element sizes.
%This verification demonstrated that the part-scale distortion induced in the AM building process can be properly predicted.
%The numerical result is consistent with the experimental measurement well, regardless of the element size.
Table \ref{tab:computationaltime1} shows the computational times for both cases using 10 Intel Xeon E5-2687W cores.
%Table 1 shows the number of nodes and elements of each mesh model.
The computational time for the 1.0 mm model is approximately 1/180$^\text{th}$ that of the 0.25 mm model.
%It can be seen that the computational time for the 1.0 mm model is about 1/180$^\text{th}$ of that for the 0.25 mm model.
\begin{table}[htbp]
\begin{center}
	\caption{Computational time for the analysis of the cantilever model with different element sizes.} 
	\includegraphics[width=8.0cm]{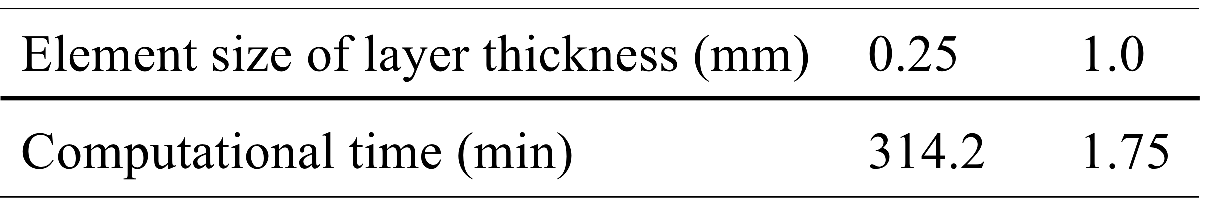}
	\label{tab:computationaltime1}
\end{center}
\end{table}

Next, to verify the validity of predicting the residual stress, we compared the experimental and numerical results of the residual stress distribution.
A cubic specimen (10mm × 10mm × 10mm) made of AlSi10Mg was fabricated under the same manufacturing conditions as for the cantilever specimen.
To measure the residual stress distribution in the building direction, the specimen was successively removed from the top surface by electrolytic polishing, and the stress was measured by X-ray diffraction (μ-X360, PULSTEC INDUSTRIAL CO., LTD.).
The measurement conditions are shown in Table \ref{tab:measurementcondition}.
The X-rays were applied to the center of the surface; the specimen was removed from the top surface to a depth of 2 mm.
%The X-rays were applied to the center of the surface.
%The specimen was removed from the top surface to a depth of 2 mm.
\begin{table}[htbp]
\begin{center}
	\caption{Measurement conditions.} 
	\includegraphics[width=8.0cm]{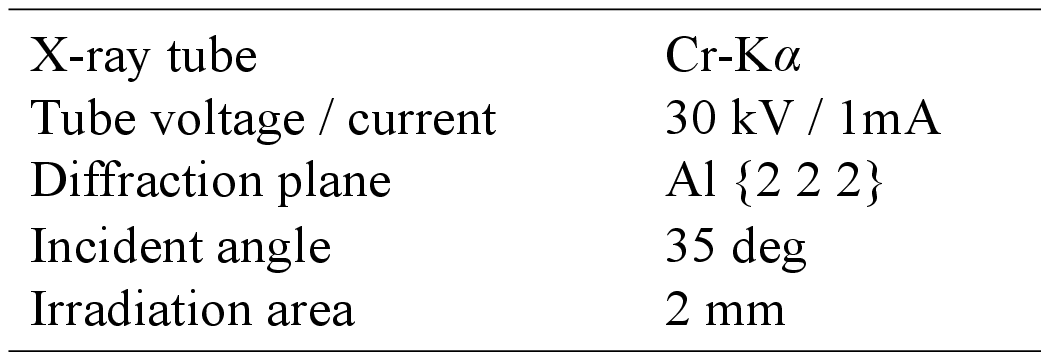}
	\label{tab:measurementcondition}
\end{center}
\end{table}
\begin{figure}[htbp]
\begin{center}
	\centering
	\subfigure[]{\includegraphics[width=5.0cm]{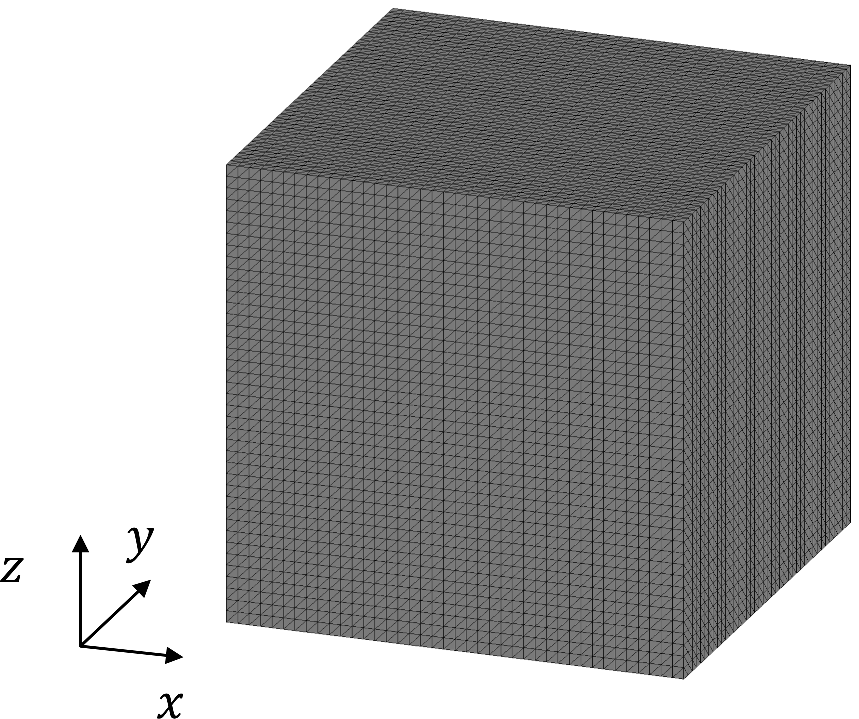}}
	\hspace{1.0cm}
	\subfigure[]{\includegraphics[width=5.0cm]{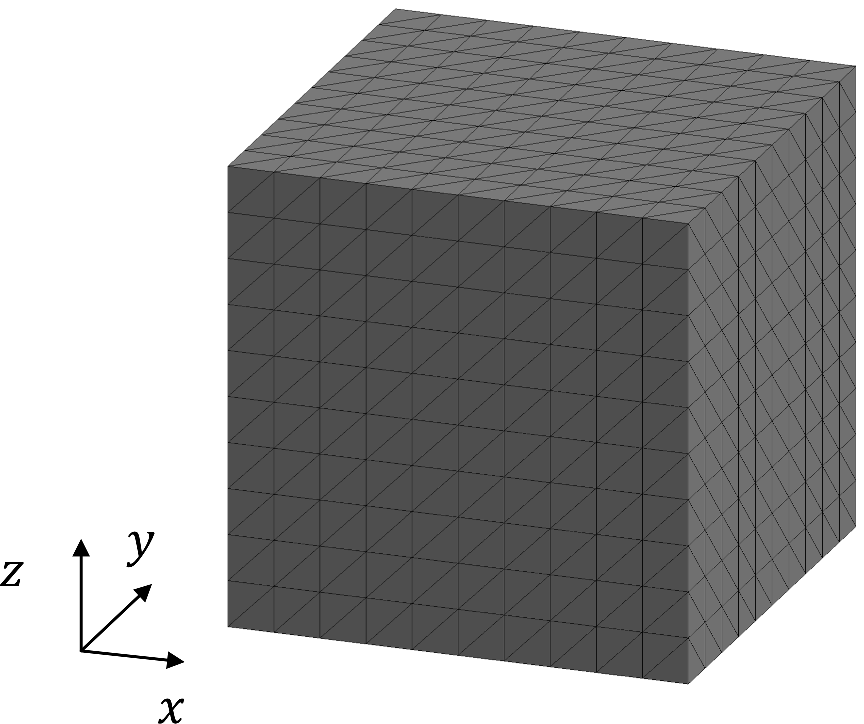}}
	\caption{Mesh model of the cubic specimen: (a) element size of 0.25 mm; (b) element size of 1.0 mm.}
	\label{fig:FemModel_cube}
\end{center}
\end{figure}
%\textcolor{blue}{Same as the validity verification of predictiong the distortion, in the numerical analysis, we prepared models with element sizes of 0.5 mm and 1.0 mm shown in Fig.\ref{fig:FemModel_cube}, and compare the accuracy and computational time.
Similar to the verification of the validity of predicting the distortion, we prepared two models with element sizes of 0.25 mm and 1.0 mm, as shown in Fig.\ref{fig:FemModel_cube}.
The meshes of the two models consist of 384,000 and 6,000 second-order tetrahedral elements, respectively.
We then evaluated the accuracy and computational time.
%	Same as the validity verification of predictiong the distortion, in the numerical analysis, models with mesh sizes of 0.5 mm and 1.0 mm shown in Fig.\ref{fig:FemModel_cube} were prepared to compare the accuracy and computational time.
The material properties and the inherent strain were set to the same values as in the analysis of the cantilever specimen.
\begin{figure}[htbp]
\begin{center}
	\includegraphics[width=10.0cm]{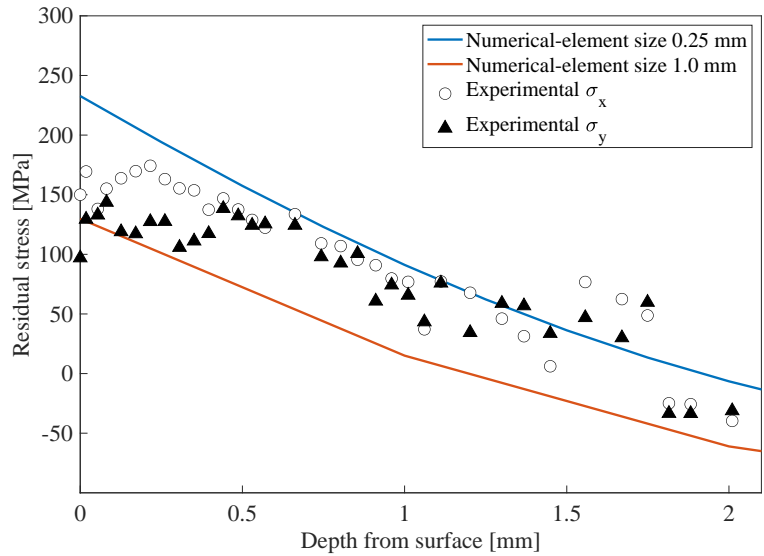}
	\caption{Comparison of the experimental and numerical results of the residual stress distribution depth profile.}
	\label{fig:StressResult}
\end{center}
\end{figure}
\begin{table}[htbp]
\begin{center}
	\caption{Computational time for the analysis of the cube model with different element sizes.} 
	\includegraphics[width=8.0cm]{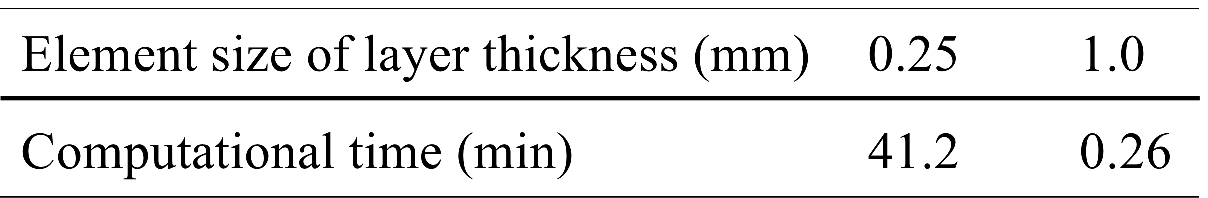}
	\label{tab:computationaltime2}
\end{center}
\end{table}

Figure \ref{fig:StressResult} shows the comparison between the numerical and experimental results in terms of the residual stress depth profile.
In both results, the maximum tensile stress was generated on the top surface that decreased gradually toward the bottom surface, where it changed to compressive stress.
This is because when the new layer solidifies and shrinks, it is restrained by the previously solidified layer, inducing a tensile force on the top surface and a compressive force on the previous layer \cite{mercelis2006residual,li2018residual}.
This verification demonstrated that the residual stress could be predicted accurately for both element sizes.
%	This verification demonstrated that the residual stress can be properly predicted by using the inherent strain obtained experimentally.
Table \ref{tab:computationaltime2} shows the computational time for both cases using 10 Intel Xeon E5-2687W cores.
%Table \ref{tab:computationaltime2} shows the computational times for both cases using 10 Intel Xeon E5-2687W cores.
The computational time for the 1.0 mm model is approximately 1/160$^\text{th}$ that of the 0.25 mm element size model.
%\textcolor{blue}{As in the verification analysis of predicting the distortion, it can be seen that the computational time for the 1.0 mm model is 1/15$^\text{th}$ of that for the 0.25 mm element size model.}
Considering the incorporation into topology optimization with iterative calculations, the element size per layer should be approximately 30 times that of the actual material layer thickness in manufacturing, i.e., the element size per layer should be 1.0 mm to reduce the computational costs in our method.
From the above, we confirmed the validity of the analytical model, and estimated the element size suitable for topology optimization.
%From the above, the validity of the proposed analytical model was confirmed.
\section{Level set-based topology optimization}\label{sec:5}
The basic idea of topology optimization is to replace the structural optimization problem with the material distribution problem, by introducing a fixed design area $D$ containing the optimal configuration and the characteristic function $\chi$, as follows
\begin{equation}
\chi  ( \bm{x} ) = \left\{ \begin{array} { l l } { 1 } & { \text { for } \bm{x} \in \mathit{\Omega} }, \\ { 0 } & { \text { for } \bm{x} \in D \backslash \mathit{\Omega} }, \end{array} \right.
\label{eq:lsf1}
\end{equation}
where $\mathit{\Omega}$ is the material domain that denotes the design domain.
The above characteristic function makes it possible to represent a configuration with an arbitrary topology.
However, it is known that topology optimization problems are commonly ill-posed \cite{allaire2002shape},
therefore, the design domain should incorporate relaxation or regularization techniques to make the problem well-posed.
The homogenization design method \cite{bendsoe1988generating} is a representative approach for relaxing the design domain.
Furthermore, level set-based shape and topology optimization methods \cite{allaire2004structural,yamada2010topology} that regularize the design space have been proposed.

In this study, we used a topology optimization method that updates the level set function, which is defined as a piecewise constant value function by a time evolution equation.
The level set function $\phi ( \bm{x} )$ takes real values between −1 and 1, and represents the boundaries $\partial \mathit{\Omega}$ between the material and void domains, using the iso-surface of $\phi ( \bm{x} )$ defined as:
\begin{equation}
\left\{ \begin{array} { l l } { 0 < \phi ( \bm{x} ) \leq 1 } & { \text { for } \bm{x} \in \mathit{\Omega} \backslash \partial \mathit{\Omega} }, \\ { \phi ( \bm{x} ) = 0 } & { \text { for } \bm{x} \in \partial \mathit{\Omega} }, \\ { - 1 \leq \phi ( \bm{x} ) < 0 } & { \text { for } \bm{x} \in D \backslash \mathit{\Omega} }. \end{array} \right.
\label{eq:lsf2}
\end{equation}
The characteristic function $\chi$ is expressed using the level set function as follows:
\begin{equation}
\tilde\chi(\phi)= \left\{ \begin{array} { l l } { 1 } & { \text { for } \phi( \bm{x} ) \geq 0 }, \\ { 0 } & { \text { for } \phi( \bm{x} ) < 0 }. \end{array} \right.
\label{eq:lsf3}
\end{equation}
We now consider a structural optimization problem that determines the optimal material distribution using the level set function.
In other words, the structural optimization problem should evaluate a level set function that minimizes the objective function, and it is described as follows:
\begin{align}
&\underset{\phi}{\text{inf}}\;\;\;F(\tilde\chi(\phi),\phi)=\int_{D}f_{d}(\bm{x})\tilde\chi(\phi)d\mathit{\Omega}+\int_{\Gamma}f_{b}(\bm{x})d\Gamma,\label{eq:lsf4}\\
&\text{subject to:}\hspace{1.5cm} \text{governing equation system},\label{eq:lsf5}
\end{align}
where $f_{d}(\bm{x})$ and $f_{b}(\bm{x})$ are the integrands of the objective function at the fixed design domain and domain boundaries, respectively.
Based on the method reported by Yamada et al. \cite{yamada2010topology}, we replace the problem of finding the optimal distribution of the level set function with a problem that solves the time evolution equation. Assuming the level set function is a function of the introduced fictitious time $t$, it is updated by the reaction-diffusion equation as follows:
\begin{equation}
\frac { \partial \phi } { \partial t } = - KF',
\label{eq:lsf7}
\end{equation}
where $K$ is a positive parameter, and $F'$ represents the topological derivative \cite{amstutz2006new}.
Because the level set function may be discontinuous everywhere in the fixed design domain, the above function is regularized by adding a Laplacian term to the second term on the right side as follows:
\begin{equation}
\frac { \partial \phi } { \partial t } = - K(F'-\tau\nabla^{2}\phi),
\label{eq:lsf8}
\end{equation}
where $\tau$ is a regularization parameter that affects the degree of diffusivity when updating the level set function, 
i.e., as $\tau$ becomes larger, the level set function leads to a smoother distribution.
Thus, by adjusting the regularization parameter $\tau$, it is possible to qualitatively adjust the geometrical complexity of the optimal configuration \cite{yamada2010topology}, which can be obtained by sequentially updating the level set function using this reaction-diffusion equation.
\section{Topology optimization considering part distortion in AM}\label{sec:6}
\subsection{Formulation of optimization problem}
We incorporate the proposed AM analytical model into the optimization framework described in the previous section.
The objective function for reducing the part distortion can be represented by minimizing the following equation:
\begin{equation}
F_{AM}=\left(\int_{\mathit{\Om}}\left|\bm{u}\right|^\be d \mathit{\Omega}\right)^{1/\be},\\
\label{eq:op1}
\end{equation}
where $\be\geq 2$ is a fixed weighting parameter.
Increasing $\be$ leads to the minimization of the maximum distortion and decreasing $\be$ leads to the minimization of the average distortion.
Furthermore, $\bm{u}$ is the sum of the distortion $\bm{u}_{i}(i=1,2,\ldots,m)$ in each active domain $\mathit{\Om_{A}}$, expressed by:
\begin{equation}
\bm{u}=\sum_{i=1}^{m}\bm{u}_{i}\hspace{5mm}\text { for } \bm{x} \in \mathit{\Omega_{A}}.\\
\end{equation}
In this study, we formulate the minimum mean compliance problem considering the part distortion in AM using the above objective function.
Consider the material domain $\mathit{\Om}$ fixed at the boundary $\mathit{\Ga_{v}}$ , with a traction $\bm{t}$ applied at $\mathit{\Ga_{t}}$.
%Let us consider the material domain $\mathit{\Om}$ that is fixed at boundary $\mathit{\Ga_{v}}$ , with a traction $\bm{t}$ applied at boundary $\mathit{\Ga_{t}}$.
The displacement field is denoted as $\bm{v} \in \mathcal{V} $ in the static equilibrium state.
The objective function of this problem is represented by minimizing the following equation
\begin{equation}
F_{MC}=\int_{\mathit{\Ga_{t}}}\bm{t}\cdot\bm{v}d\mathit{\Ga}.
\label{eq:op2}
\end{equation}
Thus, the optimization problem to determine an optimal configuration of the material domain $\mathit{\Om}$ that has the minimum mean compliance and reduces the part distortion under a volume constraint can be formulated as follows:
\begin{equation}
\begin{split}
\inf_{\tilde\chi}\hspace{17mm} &F=(1-\ga)F_{MC}+\ga F_{AM},&\\
\text{subject to}: \hspace{2mm}&G=\int_{D}\tilde\chi \hspace{1mm}d\Omega-V_\text{max}\leq 0,&\\
&-\div(\mathbb{C}\bm{\ep}(\bm{v}))=0\hspace{26mm}\text{in}\hspace{2mm}\mathit{\Om},&\\
&\bm{v}=\bm{0}\hspace{48mm}\text{on}\hspace{2mm}\mathit{\Ga_{v}},&\\
&-(\mathbb{C}\bm{\ep}(\bm{v}))\cdot\bm{n}=\bm{0}\hspace{25mm}\text{on}\hspace{2mm}\partial \mathit{\Om} \setminus \mathit{\Ga_{t}}\cup \mathit{\Ga_{v}},&\\
&-(\mathbb{C}\bm{\ep}(\bm{v}))\cdot\bm{n}=\bm{t}\hspace{25.5mm}\text{on}\hspace{2mm}\mathit{\Ga_{t}},&\\
&-\div(\mathbb{C}\bm{\ep}(\bm{u}_{i})-\mathbb{C}\bm{\ep}^{inh})=0\hspace{9mm}\text{in}\hspace{2.5mm}\mathit{\Om_{A}},&\\
&\bm{u}_{i}=\bm{0}\hspace{46.5mm}\text{on}\hspace{2mm}\mathit{\Ga_{u}},&\\
&-(\mathbb{C}\bm{\ep}(\bm{u}_{i})-\mathbb{C}\bm{\ep}^{inh})\cdot\bm{n}=\bm{0}\hspace{8.5mm}\text{on}\hspace{2mm}\partial \mathit{\Om_{A}} \setminus \mathit{\Ga_{u}},&
\label{eq:op3}
\end{split}
\end{equation}
\begin{align}
&\mathcal{V}:=\left\{\bm{v} \in H_{\Ga_{u}}^{1}(\mathit{\Om})^{N}, \hspace{1mm}  \bm{v}=\bm{0} \hspace{1mm} \text { on } \hspace{1mm} \mathit\Ga_{v}\right\},\nonumber\\
&\mathcal{U}:=\left\{\bm{u}_{i} \in H_{\Ga_{v}}^{1}(\mathit{\Om_{A}})^{N}, \hspace{1mm}  \bm{u}_{i}=\bm{0} \hspace{1mm} \text { on } \hspace{1mm} \mathit\Ga_{u}\right\},
\label{eq:op4}
\end{align}
for all indices $i = 1,2,\ldots,m$, where $0\leq\ga\leq1$ is a weighting coefficient, and $N$ is the number of spatial dimensions. 
In the above formulation, $G$ represents the volume constraint and $V_{max}$ is the upper limit of the material volume in $D$.
\subsection{Sensitivity analysis}
To derive the topological derivative of the above optimization problem, we use the adjoint variable method.
The minimum mean compliance problem is known to be a self-adjoint problem.
Therefore, the adjoint variable corresponds to the displacement field $\bm{v}$, and the topological derivative of Eq. \ref{eq:op2} is given by \cite{garreau2001topological,feijoo2005topological}
\begin{equation}
F_{MC}'=-\bm{\ep}(\bm{v}):\mathbb{A}:\bm{\ep}(\bm{v}). \label{eq:op5}\\
\end{equation}
For a plane stress problem, the constant fourth-order tensor $\mathbb{A}$ is given by:
\begin{equation}
\mathbb{A}_{i j k l}=\frac{1}{(1+\nu)^{2}}\left\{\frac{-\left(1-6 \nu+ \nu^{2}\right) E}{(1- \nu)^{2}} \delta_{i j} \delta_{k l}+2 E\left(\delta_{i k} \delta_{j l}+\delta_{i l} \delta_{j k}\right)\right\}.
\end{equation}

Next, we consider an adjoint state $\tilde{\bm{u}_{i}}\in\mathcal{U}$ for the displacement field $\bm{u}_{i}$ of the part distortion in AM.
We introduce the following adjoint system associated with the objective function Eq. \ref{eq:op1}:
\begin{align}
\begin{cases}
\hspace{2mm}-\div(\mathbb{C}\bm{\ep}(\tilde{\bm{u}_{i}}))=-\left(\int_{\mathit{\Om}} \left|\bm{u}\right|^{\be}d\mathit{\Om}\right)^{1/\be-1}\left|\bm{u}\right|^{\be-2}\bm{u}\hspace{5mm}&\text{in}\hspace{2mm}\mathit{\Om_{A}},\\
\hspace{2mm}\tilde{\bm{u}_{i}}=\bm{0}\hspace{5mm}&\text{on}\hspace{2mm}\mathit{\Ga_{u}},\\
\hspace{2mm}-(\mathbb{C}\bm{\ep}(\tilde{\bm{u}_{i}}))\cdot\bm{n}=\bm{0}\hspace{5mm}&\text{on}\hspace{2mm}\partial \mathit{\Om_{A}} \setminus \mathit{\Ga_{u}},
\label{eq:op6}
\end{cases}
\end{align}
for all indices $i = 1,2,\ldots,m$, and the topological derivative of Eq. \ref{eq:op1} is defined as follows \cite{giusti2017topology}:
\begin{equation}
F_{AM}'=\sum_{i=1}^{m}\left(-\bm{\ep}(\bm{u}_{i}):\mathbb{A}:\bm{\ep}(\tilde{\bm{u}_{i}})+\bm{\ep}^{inh}:\mathbb{A}:\bm{\ep}(\tilde{\bm{u}_{i}})\right).\label{eq:op7}
\end{equation}
\section{Numerical implementation}\label{sec:7}
\subsection{Optimization algorithm}
The optimization algorithm is as follows.
\begin{description}
\item[Step1.] The initial level set function is set.
\item[Step2.] The displacement fields $\bm{v}$ and $\bm{u}_{i}$ defined in Eqs. \ref{eq:op3} and \ref{eq:op4} are solved using FEM.
\item[Step3.] The objective function $F$ formulated using Eqs. \ref{eq:op1} and \ref{eq:op2} is calculated.
\item[Step4.] If the objective function converges, the optimization procedure is terminated; otherwise, the adjoint field $\tilde{\bm{u}_{i}}$ defined in Eq. \ref{eq:op6} is solved using FEM, and the topological derivatives with respect to the objective function are calculated using Eqs. \ref{eq:op5} and \ref{eq:op7}.
\item[Step5.] The level set function is updated using the time evolution equation given by Eq. \ref{eq:lsf8}, and then, the optimization procedure returns to the second step.
\end{description}
\subsection{Numerical scheme for the governing equation}
The finite elements are generated during the iterative optimization procedure.
From the perspective of computational cost, we use the ersatz material approach \cite{allaire2004structural}. 
We assume that the void domain is a structural material with a relatively small Young's modulus.
In addition, we assume that the boundary between the material and void domains has a smoothly distributed material property; i.e., we use the extended elastic tensor $\tilde{\mathbb{C}}$ to solve the governing equations Eqs. \ref{eq:op3} and \ref{eq:op4} in the fixed design domain $D$ as follows:
\begin{equation}
\tilde{\mathbb{C}}(\phi;w) = \left\{(1-d)H(\phi;w)+d\right\}\mathbb{C},\label{eq:ni1}
\end{equation}
where $H(\phi;w)$ is defined as:
\begin{equation}
H(\phi;w) := \left\{\begin{array}{ll}
1 & \text { for } \phi>w, \\
\frac{1}{2}+\frac{\phi}{w}\left(\frac{15}{16}-\frac{\phi^{2}}{w^{2}}\left(\frac{5}{8}-\frac{3}{16} \frac{\phi^{2}}{w^{2}}\right)\right) & \text { for }-w \leq \phi \leq w, \\
0 & \text { for } \phi<-w,
\end{array}\right.\label{eq:ni2}
\end{equation}
with $w$ representing the width of the transition, and $d$ the ratio of the Young's modulus for the structural and void materials.
%where $w$ represents the width of the transition, and $d$ represents the ratio of the Young's modulus for the structural and void materials.
\subsection{Normalization for topological derivative scaling}
The scale of the topological derivative $F_{MC}'$ and $F_{AM}'$ is significantly affected by the fixed design domain scale and boundary condition settings, therefore, we used a normalized topological derivative $ \tilde{F}'$ defined by:
%The scale of the topological derivative $F_{MC}'$ and $F_{AM}'$ is significantly affected by the fixed design domain scale and boundary condition settings.
%Therefore, we use a normalized topological derivative $ \tilde{F}'$ defined by
\begin{align}
\tilde{F}'= (1-\ga)\dfrac{ F_{MC}'\int_{D}d\mathit\Omega}{\int_{D}| F_{MC}'|d\mathit\Omega}+\ga\dfrac{ F_{AM}'\int_{D}d\mathit\Omega}{\int_{D}| F_{AM}'|d\mathit\Omega}.\label{eq:ni3}
\end{align}
\section{Numerical examples}\label{sec:8}
\begin{figure}[htbp]
\begin{center}
	\centering
	\subfigure[]{\includegraphics[width=9.5cm]{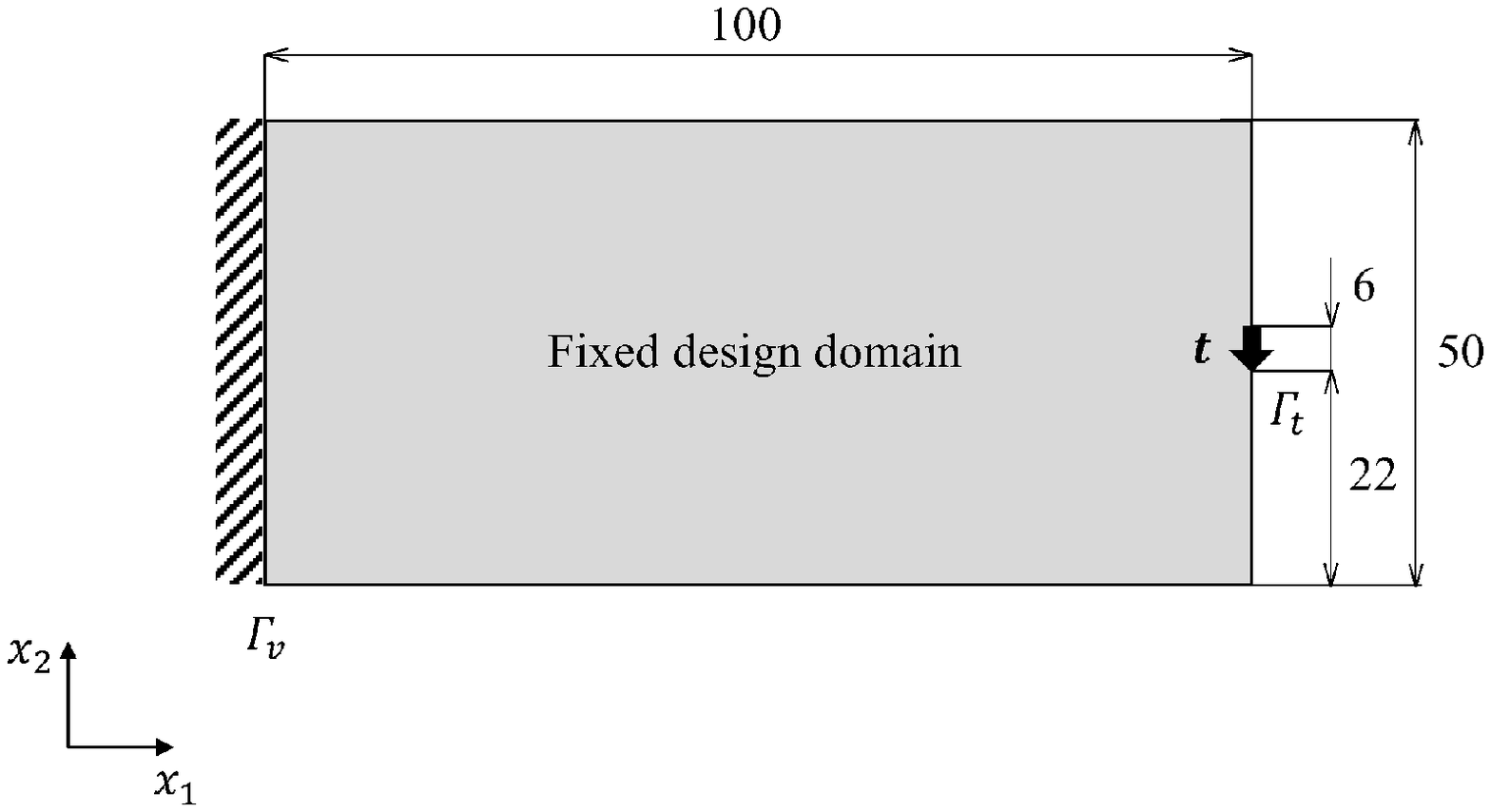}}
	%				\hspace{0.1cm}
	\subfigure[]{\includegraphics[width=9.5cm]{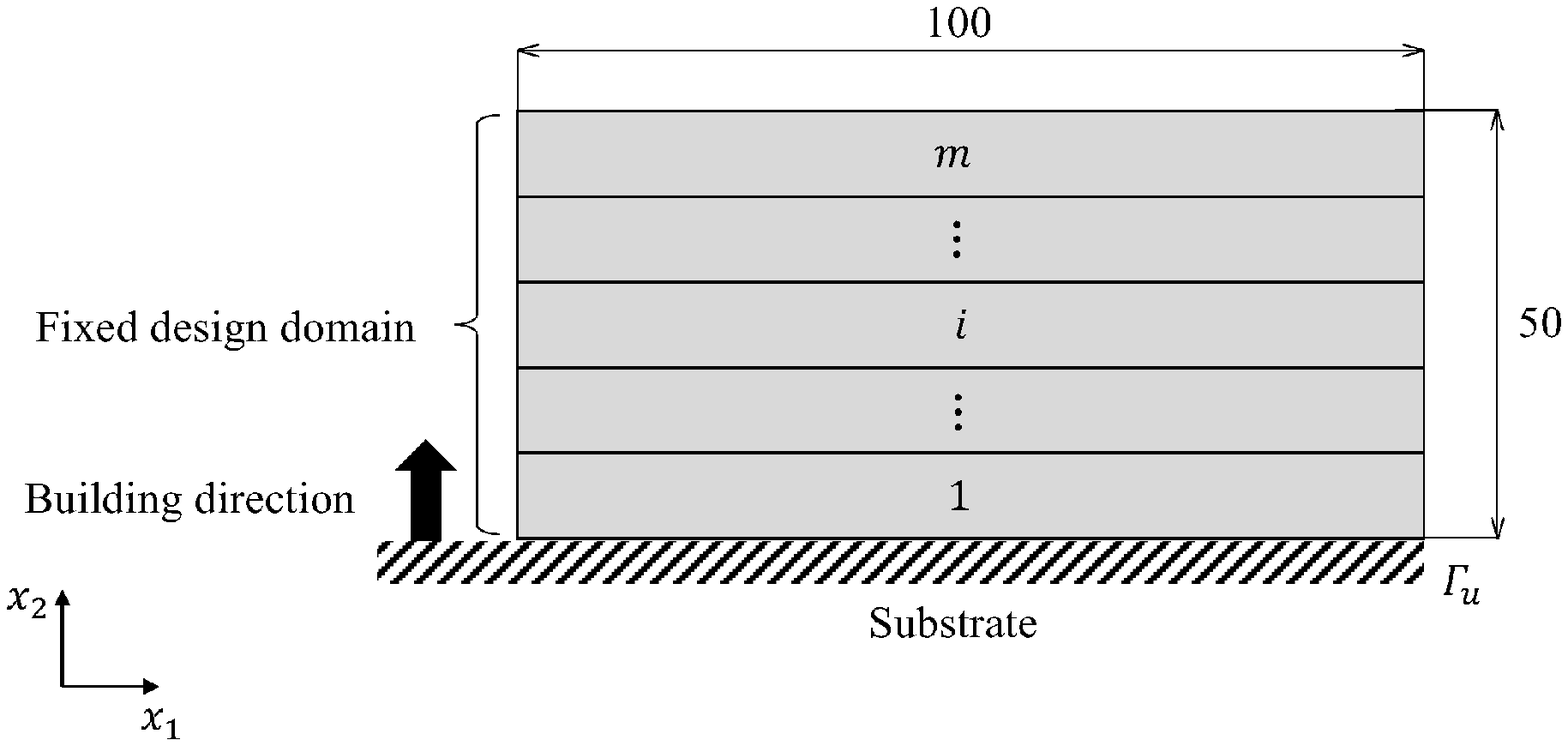}}
	\caption{Fixed design domain and boundary conditions for the cantilever model: (a) minimum mean compliance problem and (b) mechanics problem of AM.}
	\label{fig:NumericalExampleCanti}
\end{center}
\end{figure}

\begin{figure}[htbp]
\begin{center}
	\centering
	\subfigure[]{\includegraphics[width=11cm]{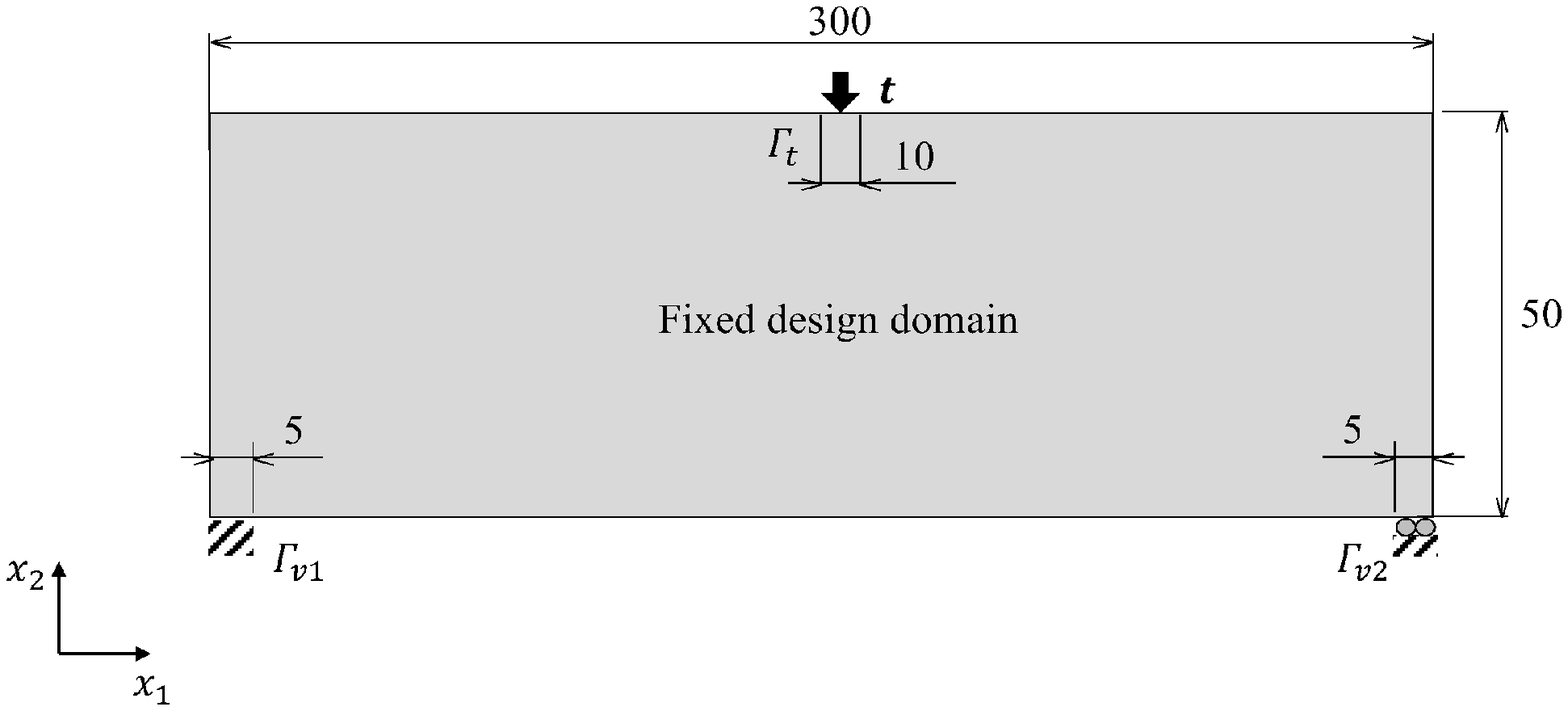}}
	%				\hspace{0.1cm}
	\subfigure[]{\includegraphics[width=11cm]{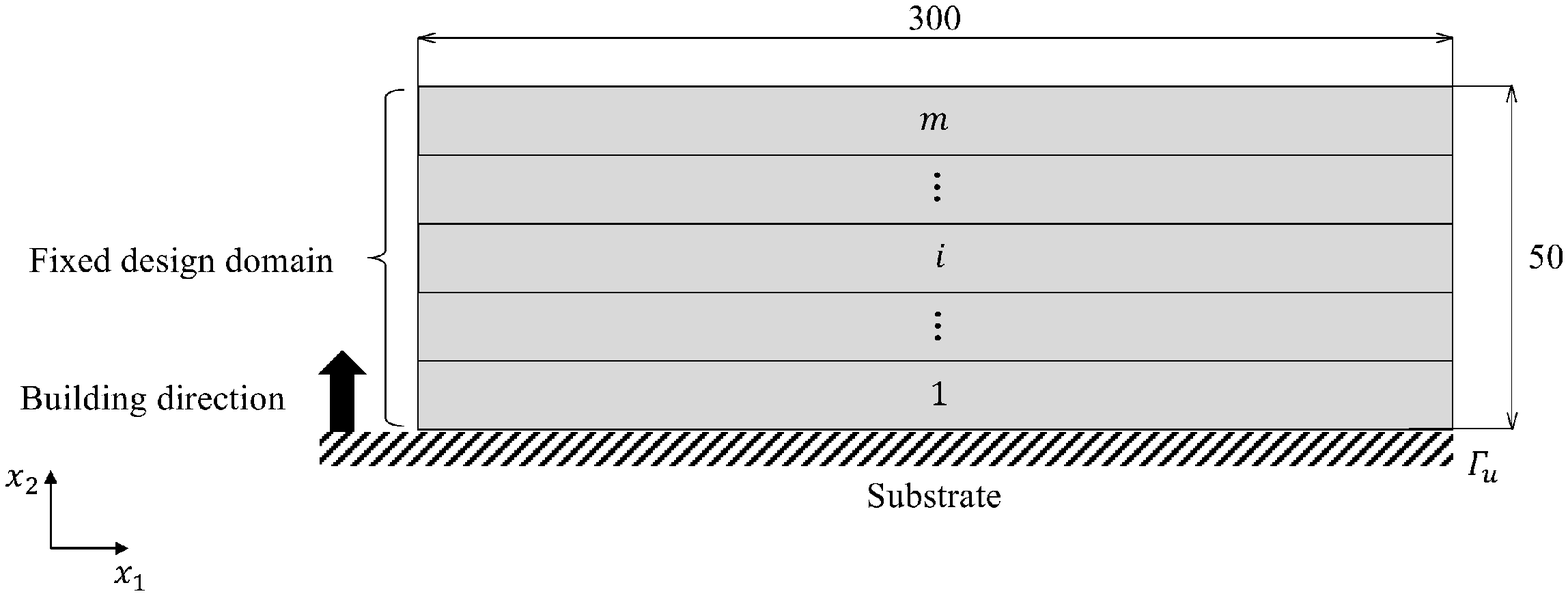}}
	\caption{Fixed design domain and boundary conditions for the Messerschmitt-Bolkow-Blohm (MBB) beam model: (a) minimum mean compliance problem and (b) mechanics problem of AM.}
	\label{fig:NumericalExampleMBB}
\end{center}
\end{figure}

\sloppy In this section, numerical examples are presented to demonstrate the effectiveness and validity of the proposed optimization method for two-dimensional minimum compliance problems that consider the part distortion in AM.
%In this section, numerical examples are presented to demonstrate the effectiveness and validity of the proposed optimization method for two-dimensional minimum compliance problems, considering the part distortion in AM.
The fixed design domain and boundary conditions of the two models are shown in Figs. \ref{fig:NumericalExampleCanti} and \ref{fig:NumericalExampleMBB}, respectively.
The fixed design domain in Figs. \ref{fig:NumericalExampleCanti}(b) and \ref{fig:NumericalExampleMBB}(b) are divided into $m = 50$ layers with a layer thickness of 1 in the building direction.
The elastic material has a Young's modulus of 75 GPa, and Poisson's ratio of 0.34.
The applied traction $\bm{t}$ was set to 10 and the inherent strain component was set to the value in Eq. \ref{eq:am13}.
The upper limit of the allowable volume was set to 50\% of the volume of the fixed design domain.
%The regularization parameter $\tau$ was set to 1 $\times$ $10^{-4}$ for the cantilever model and 1 $\times$ $10^{-3}$ for the MBB beam model.
The parameter $K$ in Eq. \ref{eq:lsf8} was set to 0.8;
the weighting parameter $\be$ in Eq. \ref{eq:op3} was set to 5.
%The parameter $K$ in Eq. \ref{eq:lsf8} was set to 0.8.
%The weighting parameter $\be$ in Eq. \ref{eq:op3} was set to 5.
The parameters $w$ and $d$ in Eq. \ref{eq:ni2} were set to 0.5 and 1 $\times$ $10^{-3}$, respectively.
We examined the dependency of the optimal configurations with respect to the different settings of the regularization parameter $\tau$ and weighting coefficient $\ga$.

\begin{figure}[htbp]
\begin{center}
	\centering
	\subfigure[]{\includegraphics[width=4.0cm]{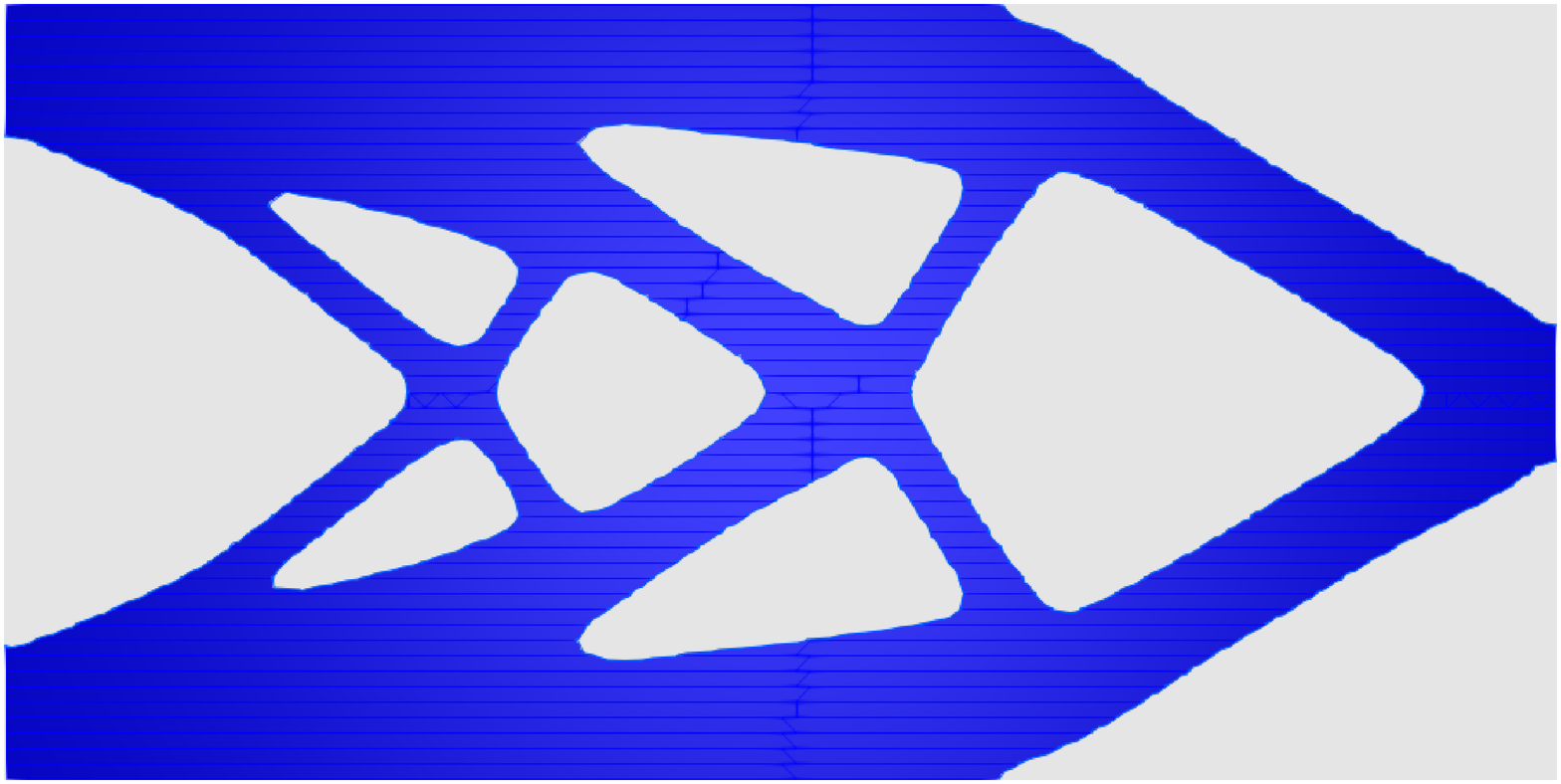}}
	%				\hspace{0.1cm}
	\subfigure[]{\includegraphics[width=4.0cm]{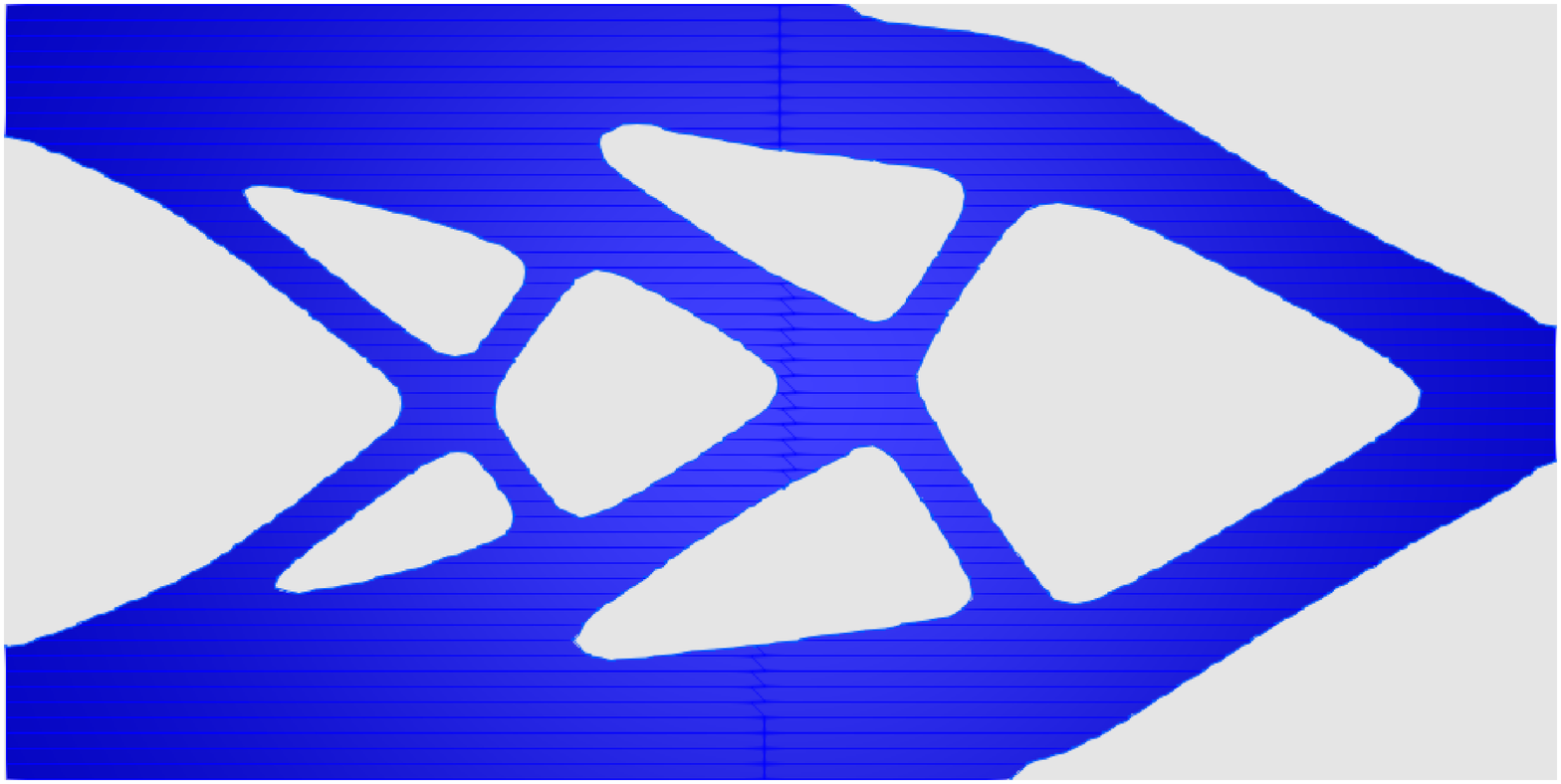}}
	%				\hspace{0.1cm}
	\subfigure[]{\includegraphics[width=4.0cm]{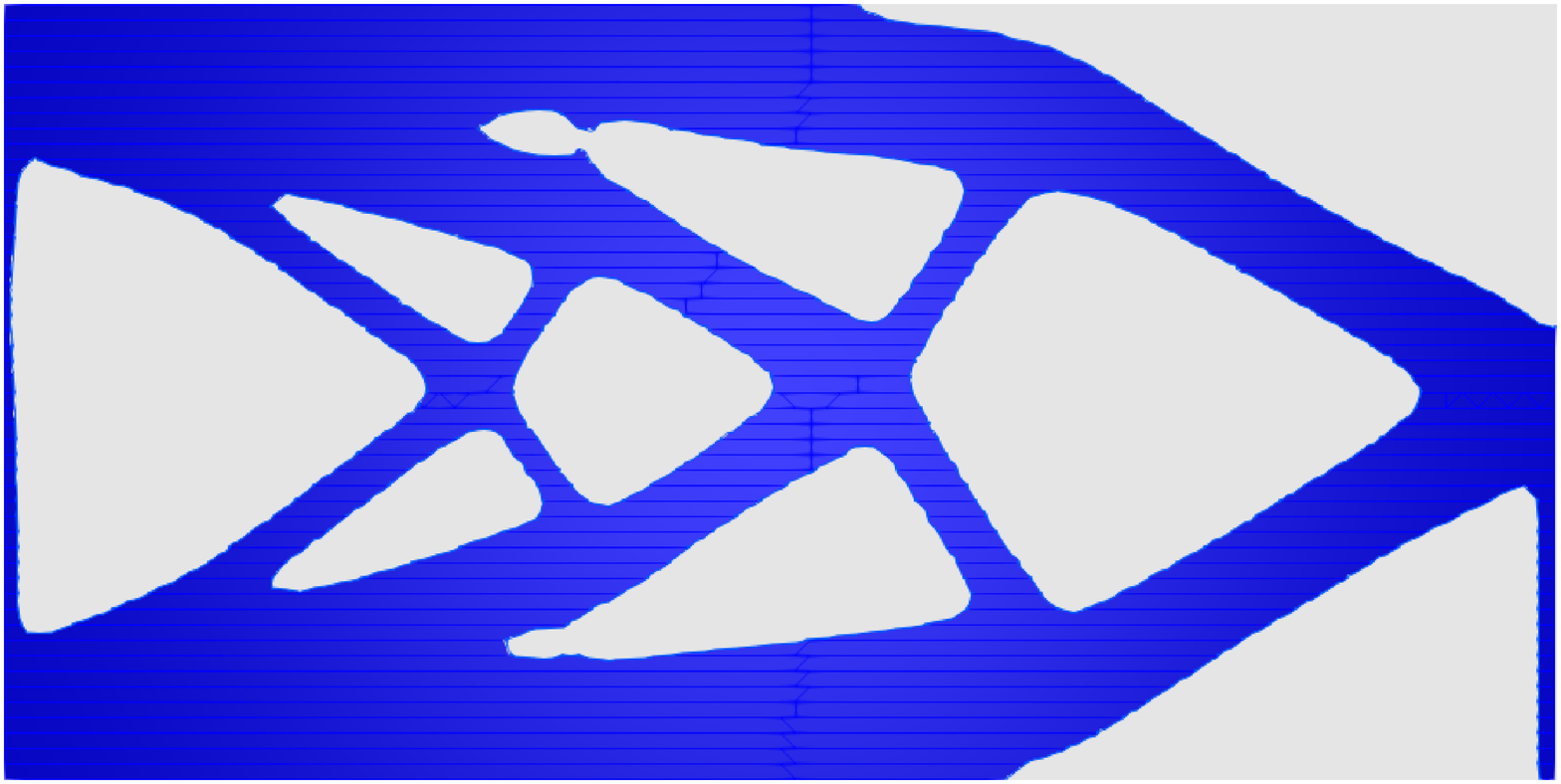}}
	%				\hspace{0.1cm}
	\subfigure[]{\includegraphics[width=4.0cm]{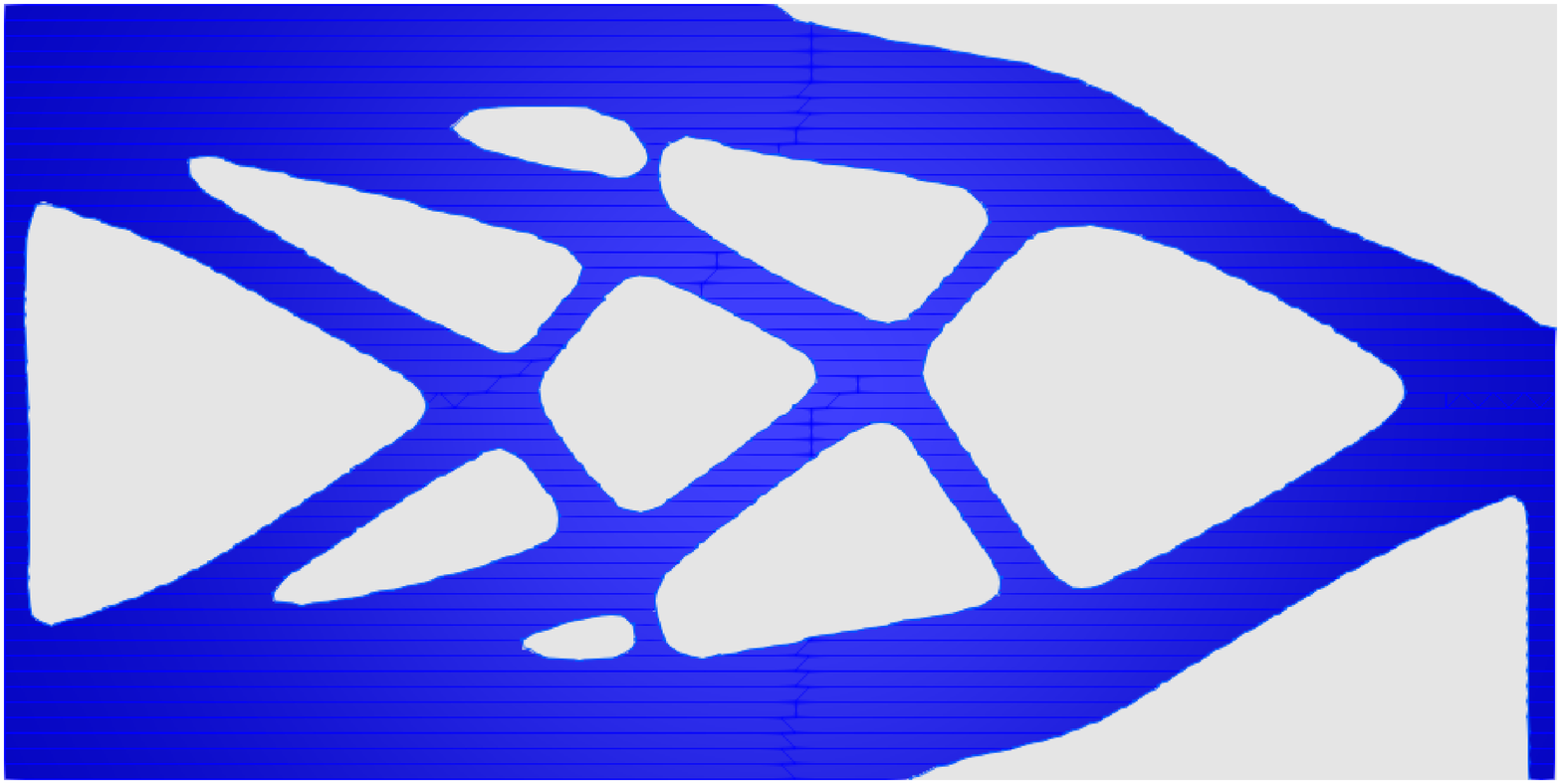}}
	%				\hspace{0.1cm}
	\subfigure[]{\includegraphics[width=4.0cm]{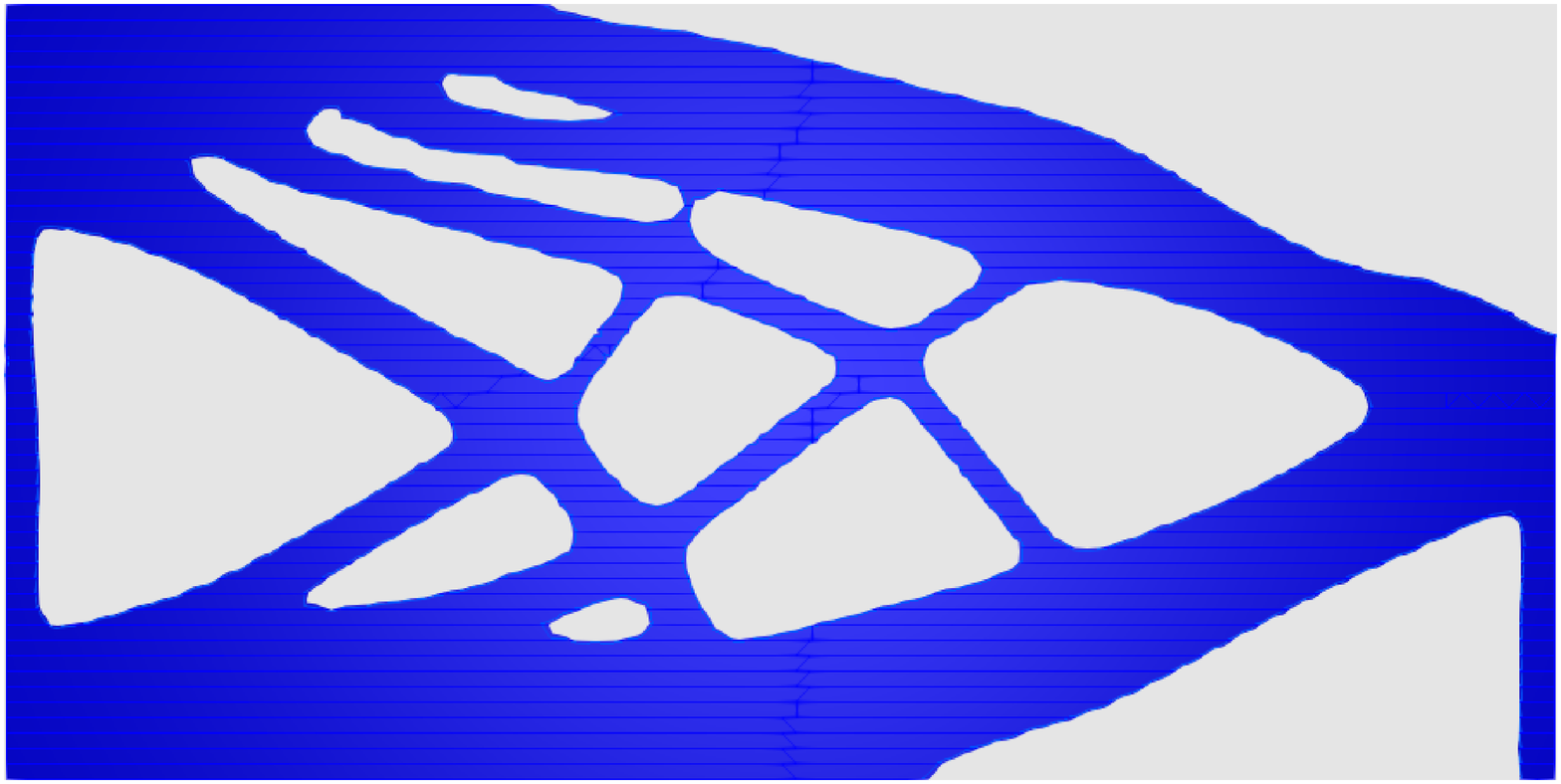}}
	%				\hspace{0.1cm}
	\subfigure[]{\includegraphics[width=4.0cm]{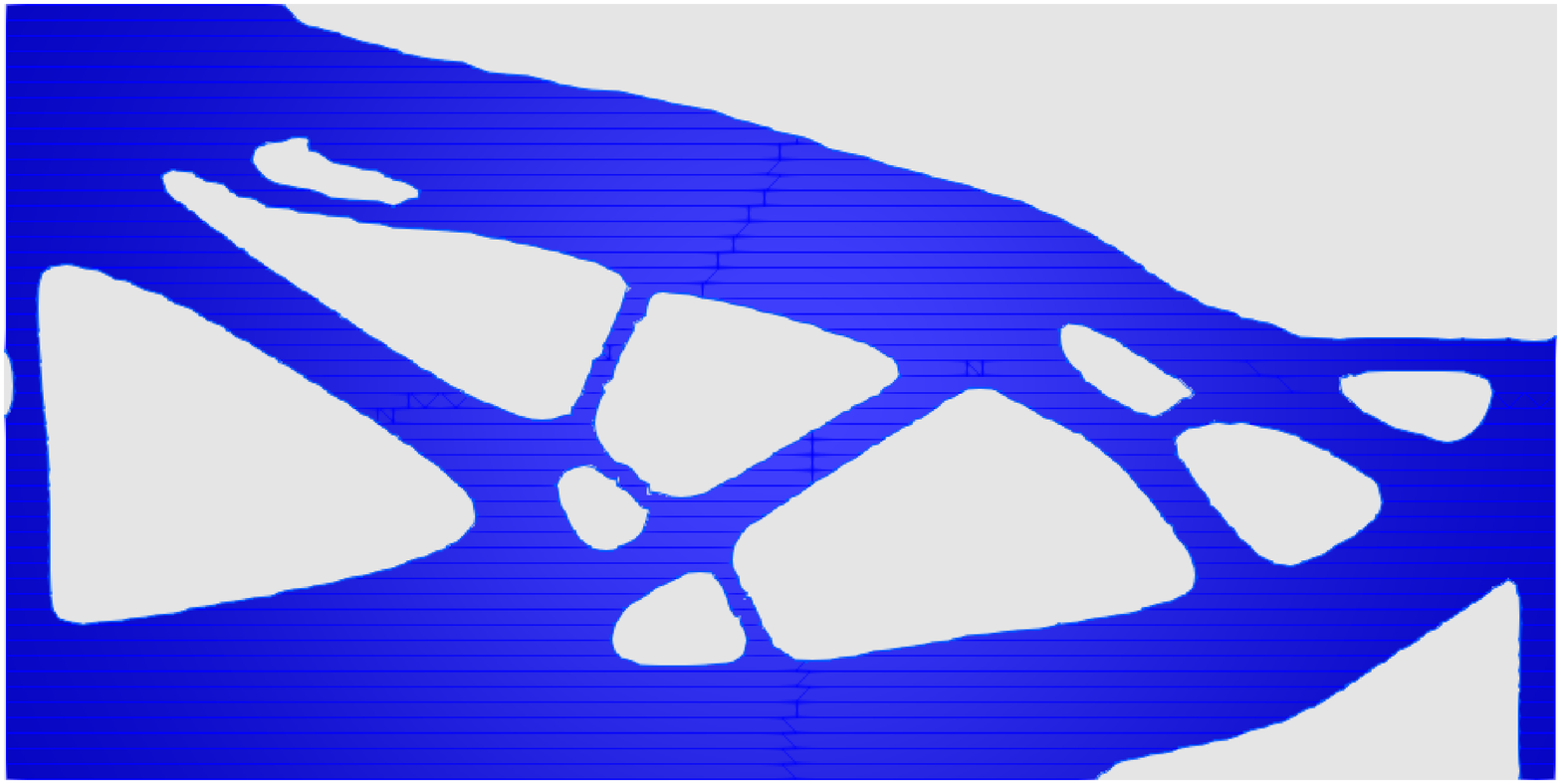}}
	\caption{Optimal configurations for the cantilever model $\tau = 1\times10^{-4}$: (a)$\ga = 0$; (b)$\ga = 0.03$; (c)$ \ga = 0.05$; (d)$\ga = 0.10$; (e)$\ga = 0.15$; (f)$\ga = 0.20$. }
	\label{fig:resultga_Canti_t1e4}
\end{center}
\end{figure}
\begin{figure}[htbp]
\begin{center}
	\centering
	\subfigure[]{\includegraphics[width=4.0cm]{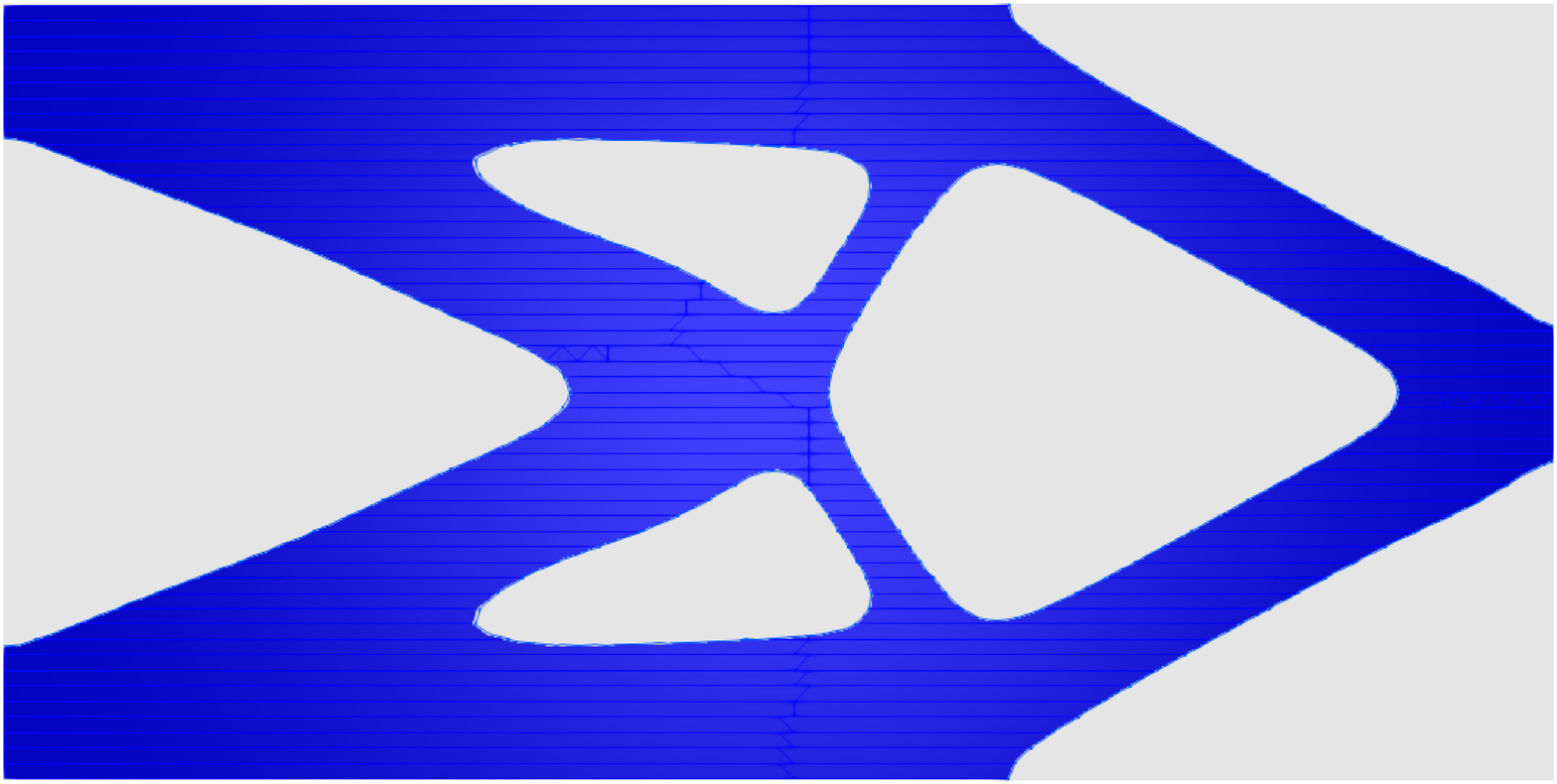}}
	%				\hspace{0.1cm}
	\subfigure[]{\includegraphics[width=4.0cm]{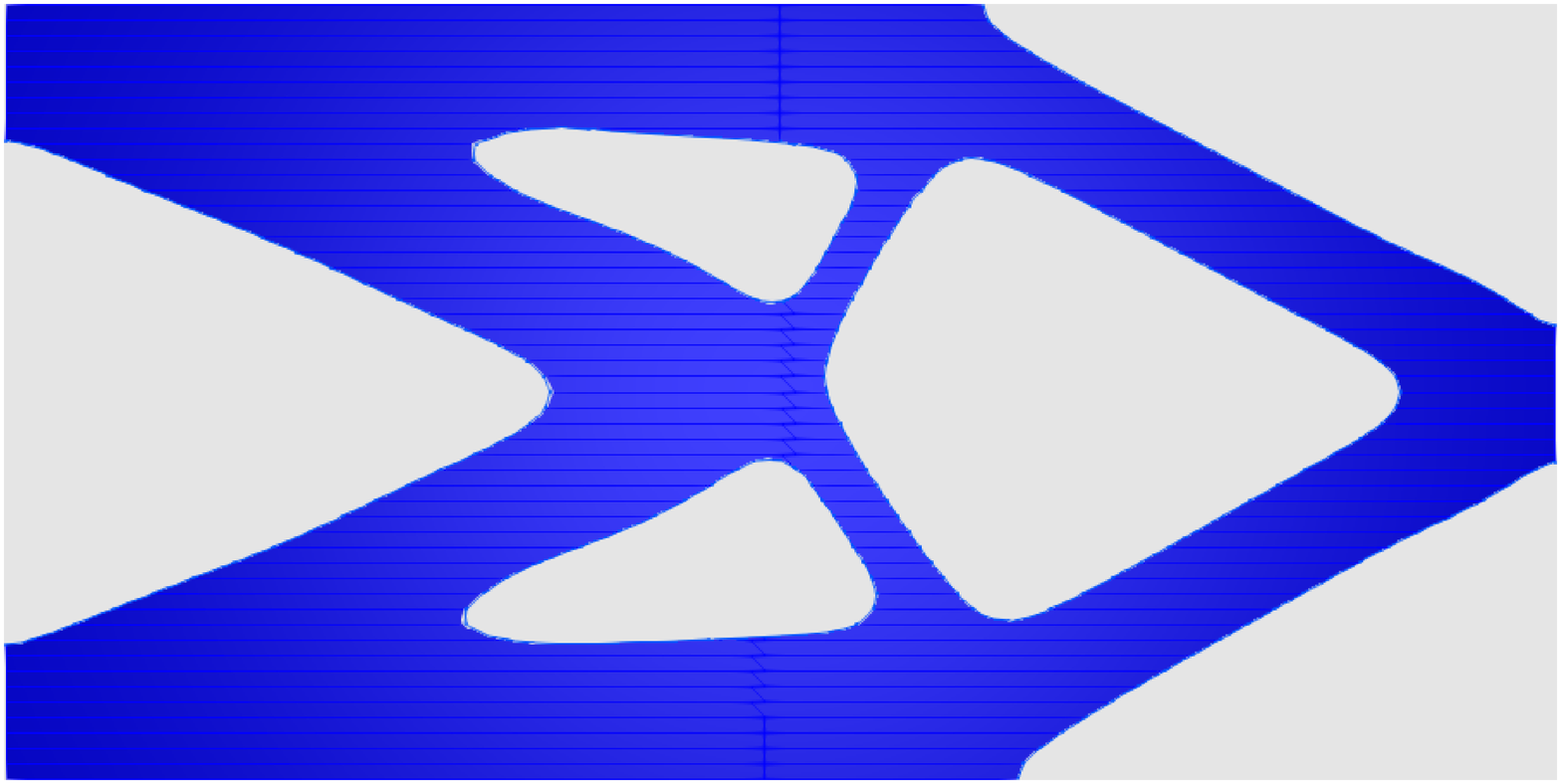}}
	%				\hspace{0.1cm}
	\subfigure[]{\includegraphics[width=4.0cm]{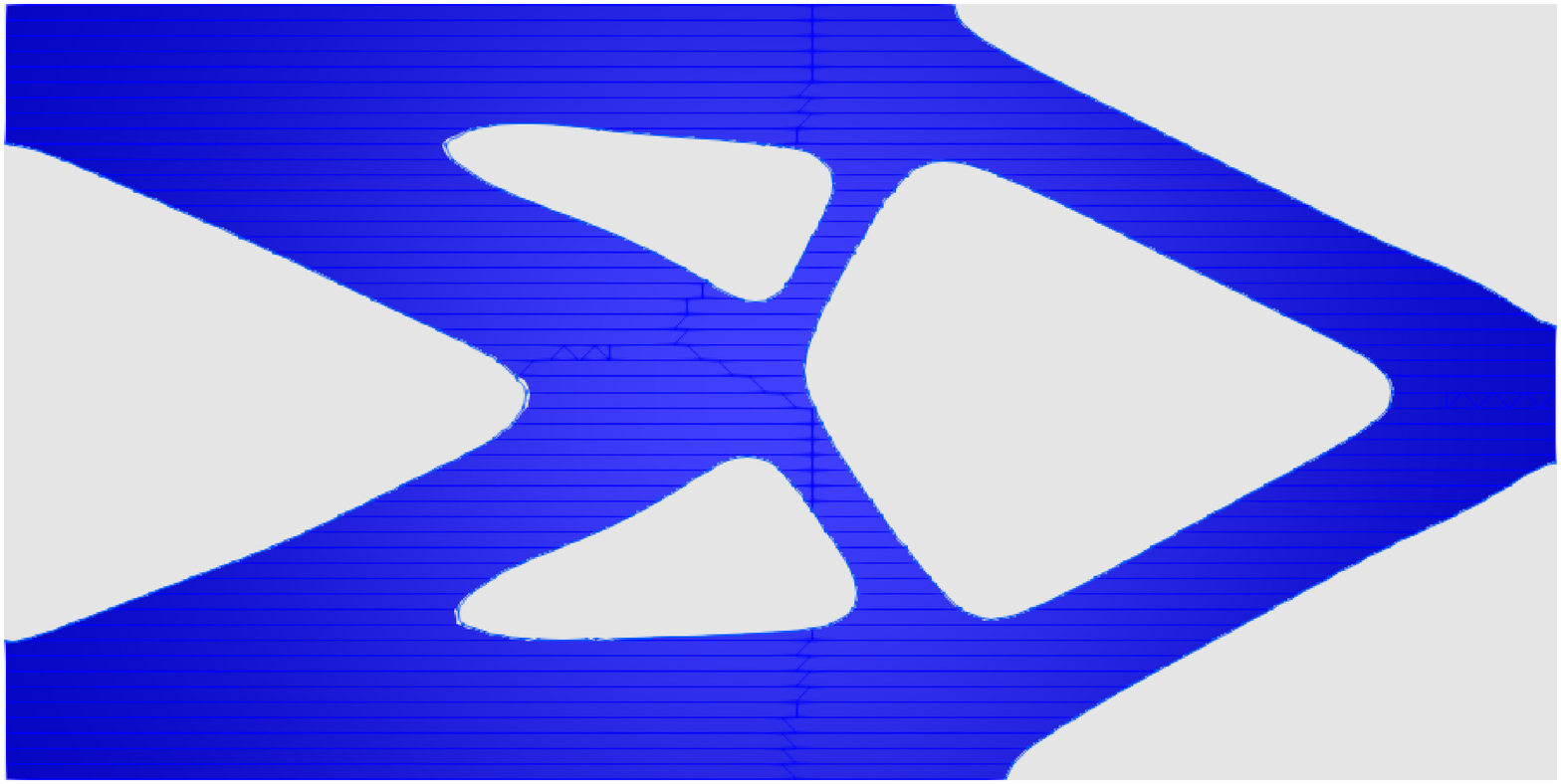}}
	%				\hspace{0.1cm}
	\subfigure[]{\includegraphics[width=4.0cm]{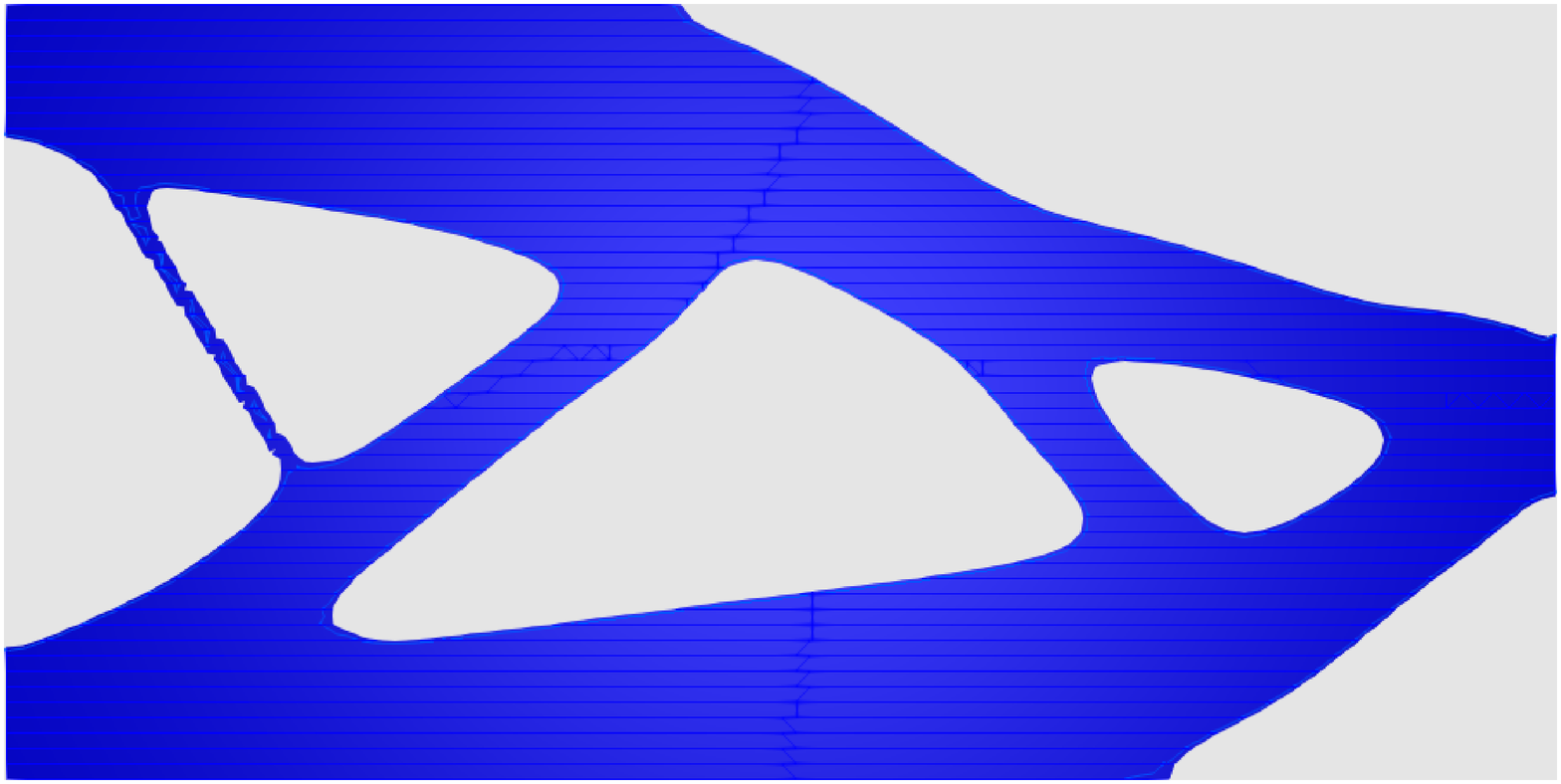}}
	%				\hspace{0.1cm}
	\subfigure[]{\includegraphics[width=4.0cm]{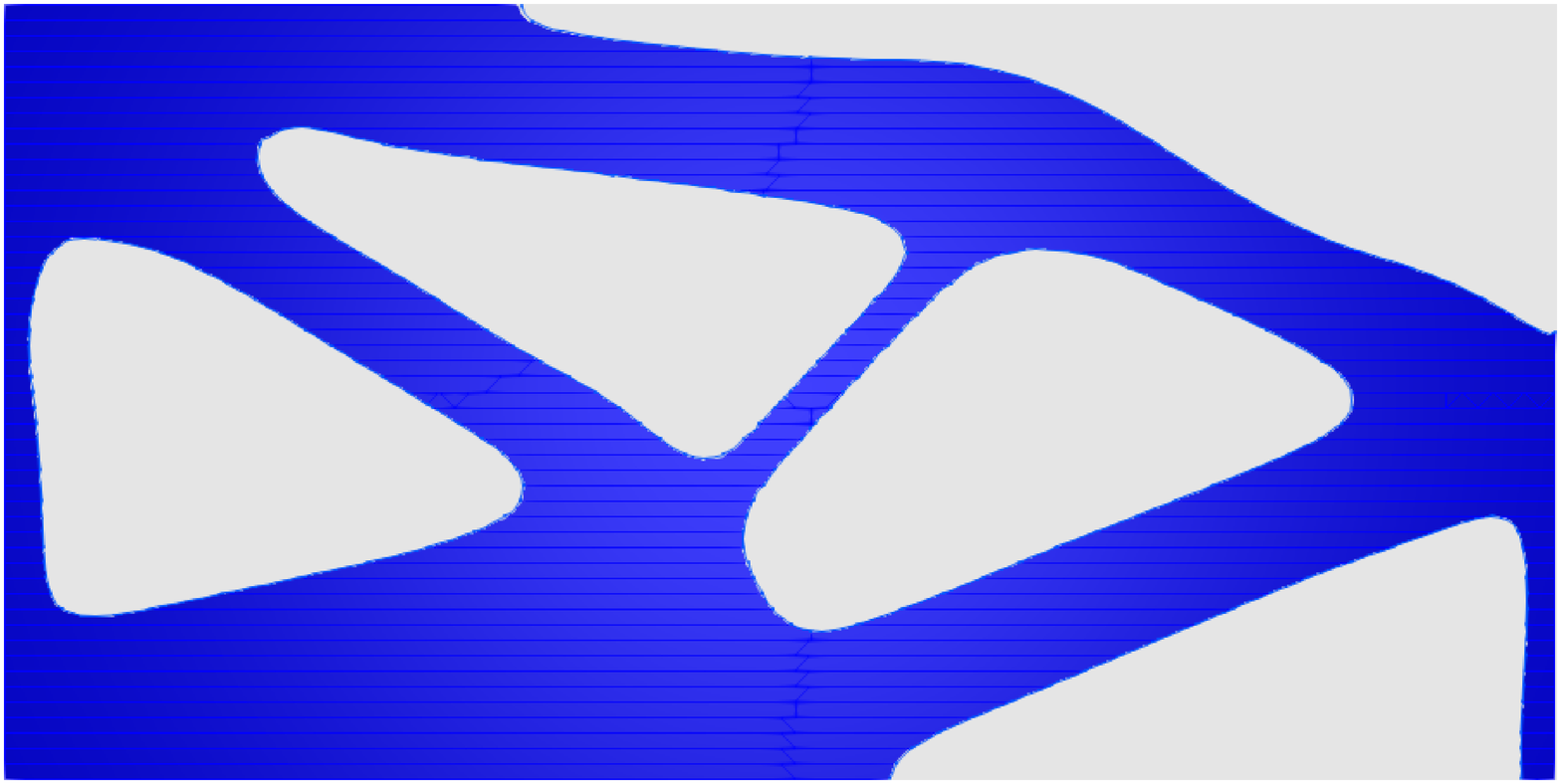}}
	%				\hspace{0.1cm}
	\subfigure[]{\includegraphics[width=4.0cm]{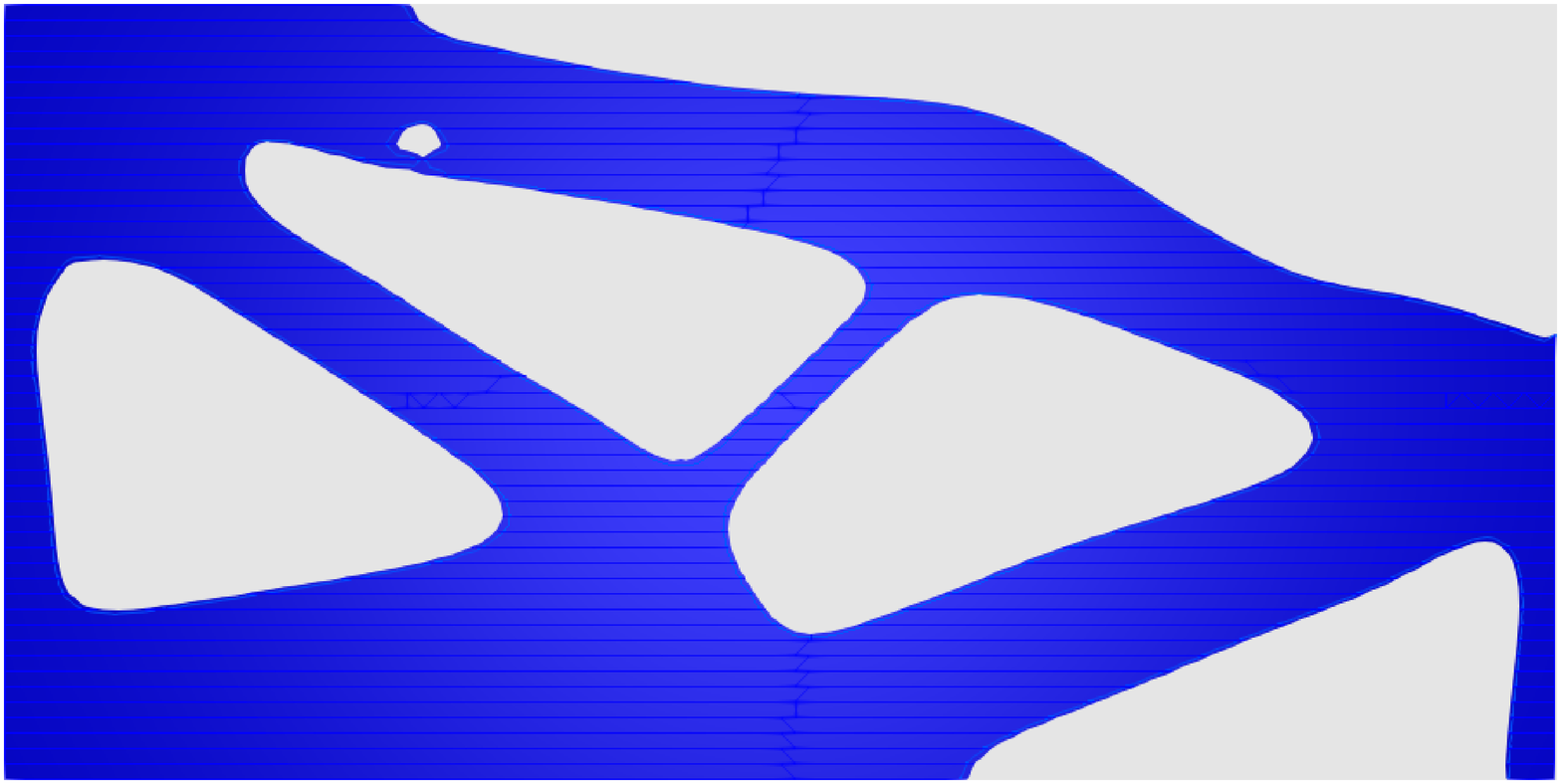}}
	\caption{Optimal configurations for the cantilever model $\tau = 1\times10^{-3}$: (a)$\ga = 0$; (b)$\ga = 0.03$; (c)$ \ga = 0.05$; (d)$\ga = 0.10$; (e)$\ga = 0.15$; (f)$\ga = 0.20$. }
	\label{fig:resultga_Canti_t1e3}
\end{center}
\end{figure}
\begin{figure}[htbp]
\begin{center}
	\centering
	\subfigure[]{\includegraphics[width=6.0cm]{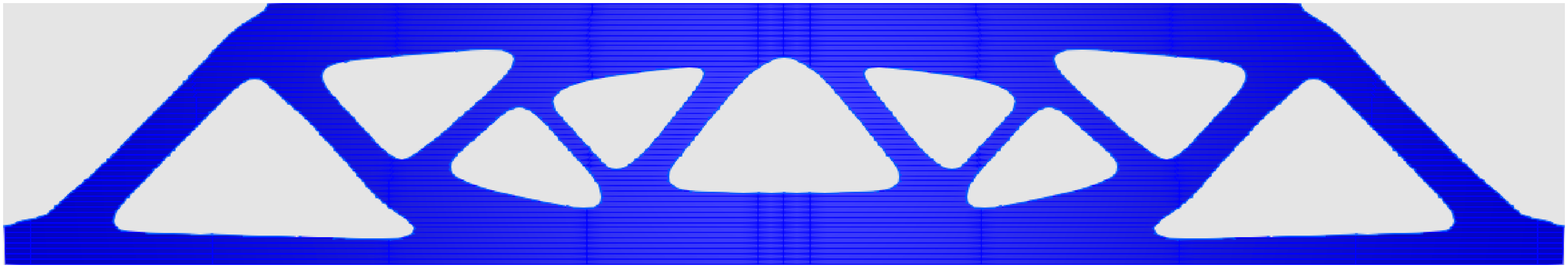}}
	%				\hspace{0.1cm}
	\subfigure[]{\includegraphics[width=6.0cm]{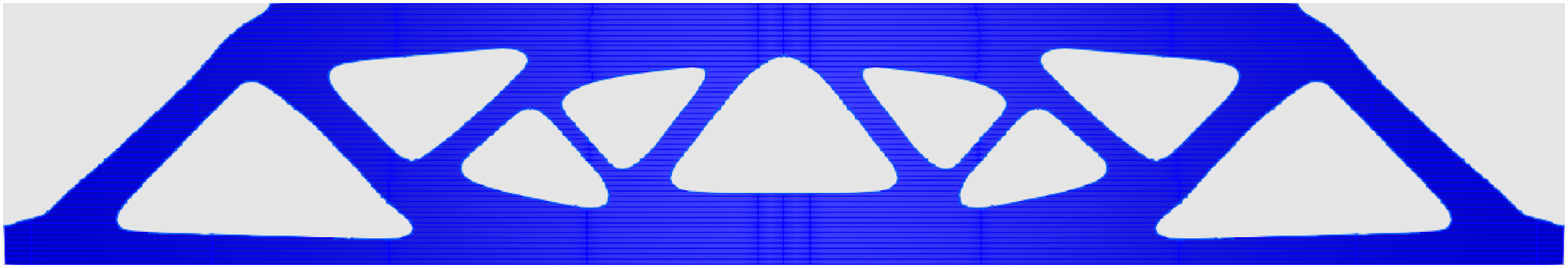}}
	%				\hspace{0.1cm}
	\subfigure[]{\includegraphics[width=6.0cm]{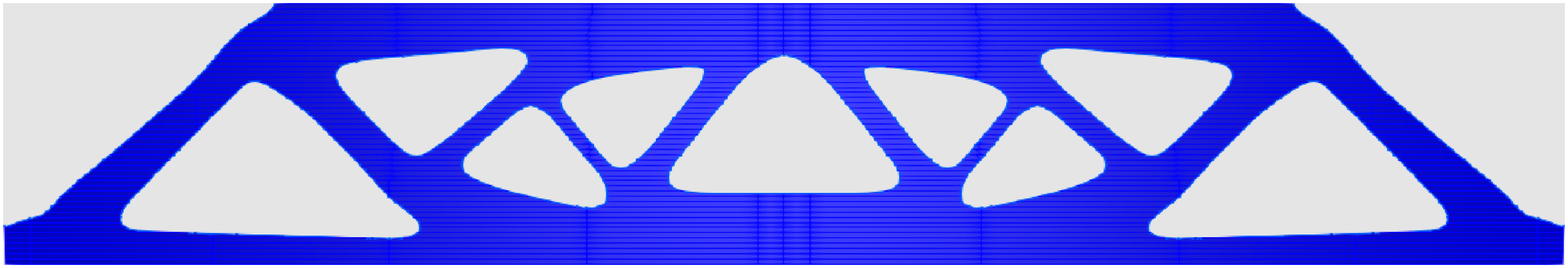}}
	%				\hspace{0.1cm}
	\subfigure[]{\includegraphics[width=6.0cm]{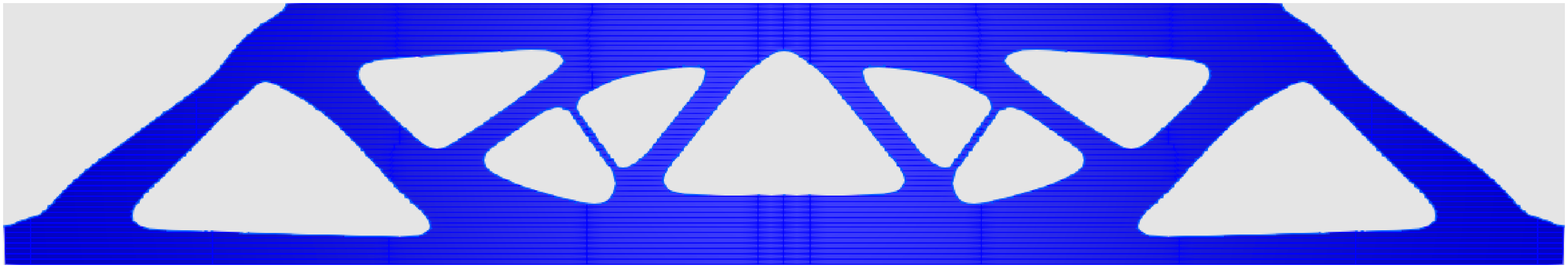}}
	%				\hspace{0.1cm}
	\subfigure[]{\includegraphics[width=6.0cm]{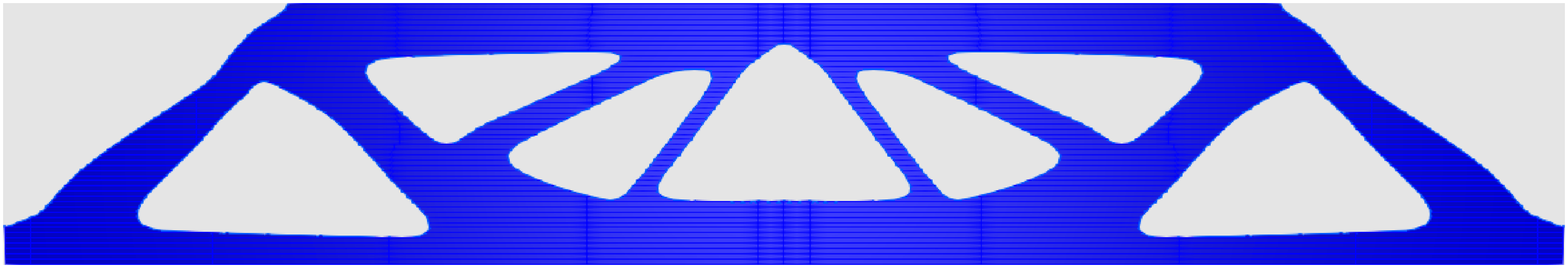}}
	%				\hspace{0.1cm}
	\subfigure[]{\includegraphics[width=6.0cm]{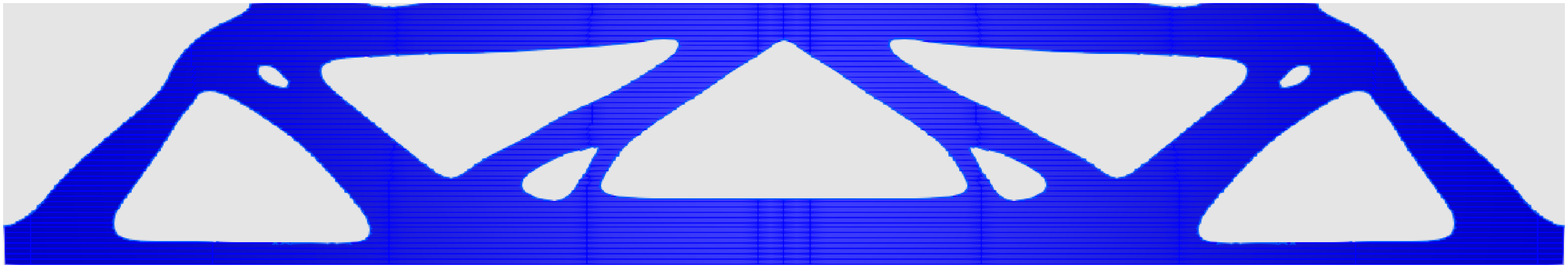}}
	\caption{Optimal configurations for the MBB beam model $\tau = 1\times10^{-4}$: (a)$\ga = 0$; (b)$\ga = 0.03$; (c)$ \ga = 0.05$; (d)$\ga = 0.10$; (e)$\ga = 0.15$; (f)$\ga = 0.20$. }
	\label{fig:resultga_MBB_t1e4}
\end{center}
\end{figure}
\begin{figure}[htbp]
\begin{center}
	\centering
	\subfigure[]{\includegraphics[width=6.0cm]{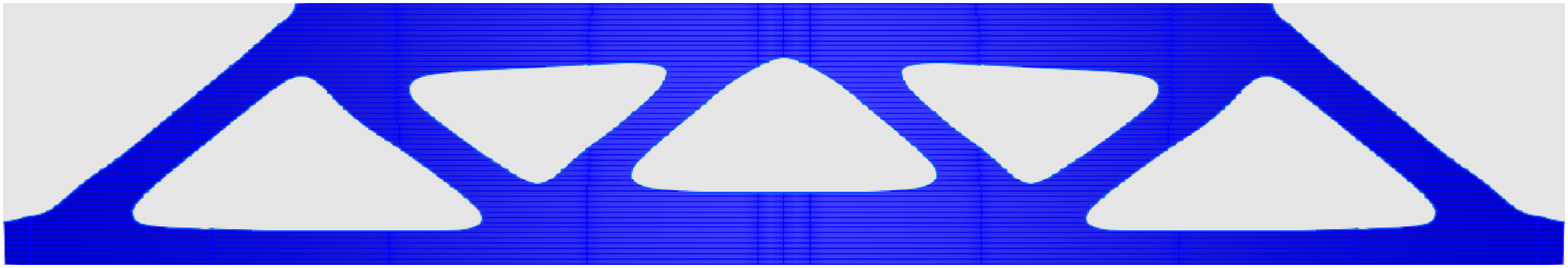}}
	%				\hspace{0.1cm}
	\subfigure[]{\includegraphics[width=6.0cm]{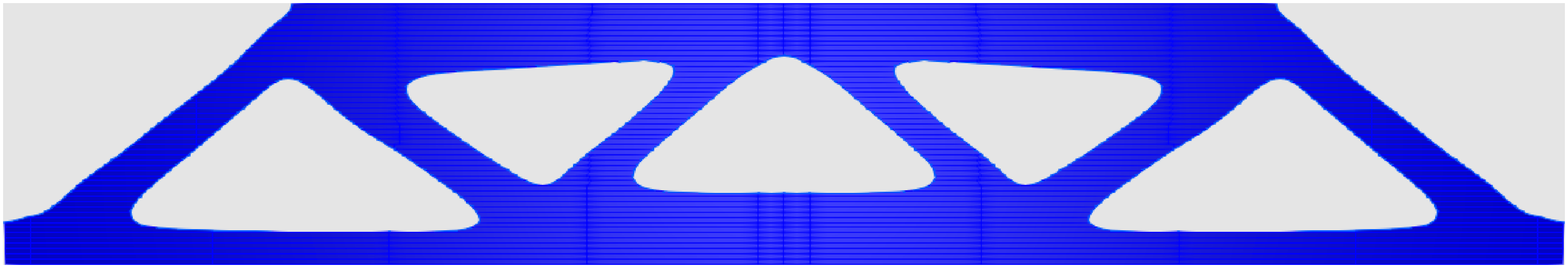}}
	%				\hspace{0.1cm}
	\subfigure[]{\includegraphics[width=6.0cm]{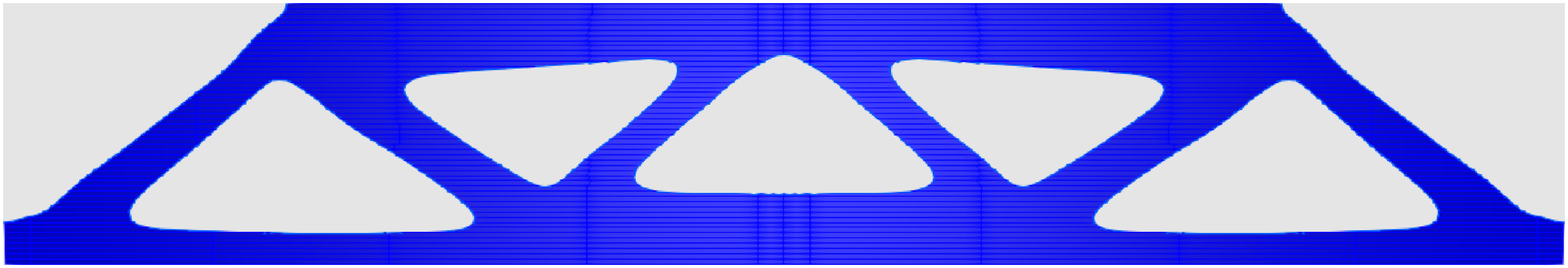}}
	%				\hspace{0.1cm}
	\subfigure[]{\includegraphics[width=6.0cm]{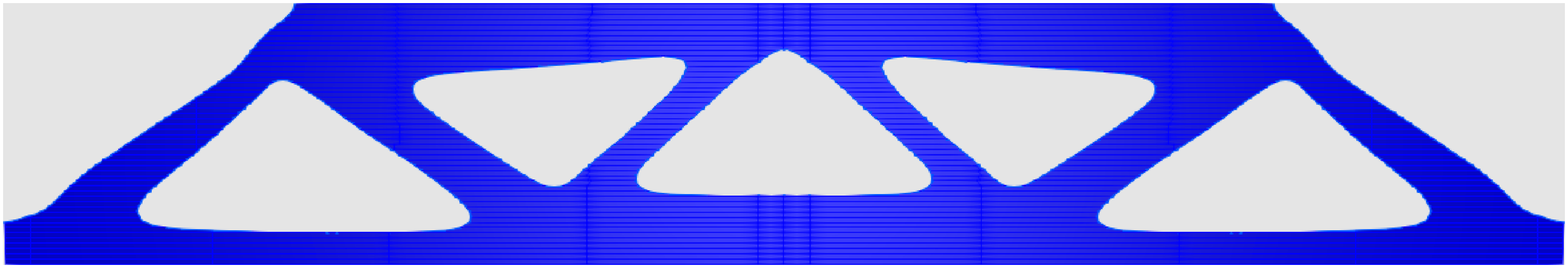}}
	%				\hspace{0.1cm}
	\subfigure[]{\includegraphics[width=6.0cm]{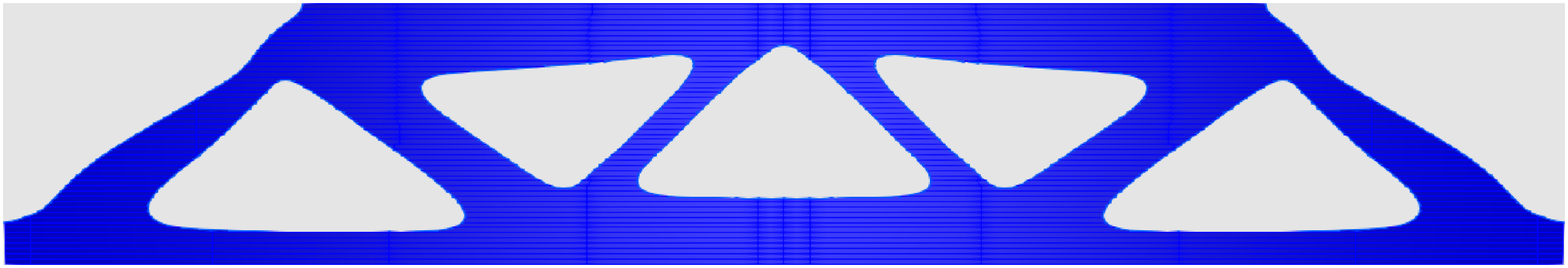}}
	%				\hspace{0.1cm}
	\subfigure[]{\includegraphics[width=6.0cm]{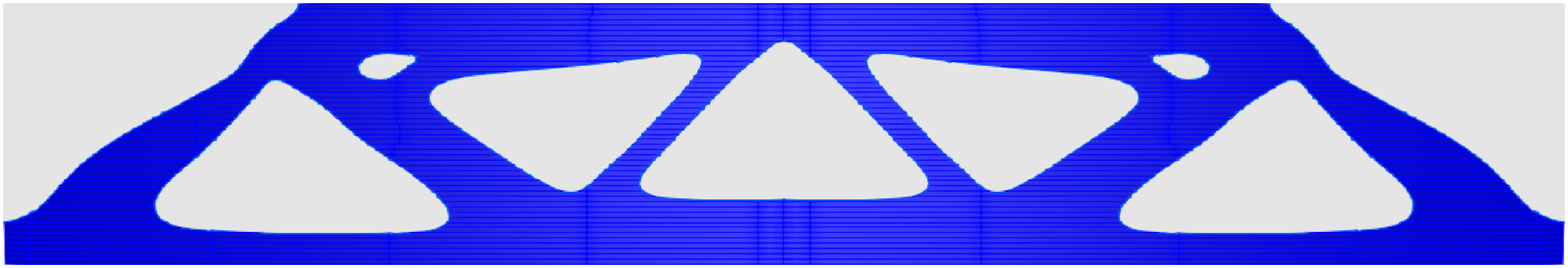}}
	\caption{Optimal configurations for the MBB beam model $\tau = 1\times10^{-3}$: (a)$\ga = 0$; (b)$\ga = 0.03$; (c)$ \ga = 0.05$; (d)$\ga = 0.10$; (e)$\ga = 0.15$; (f)$\ga = 0.20$. }
	\label{fig:resultga_MBB_t1e3}
\end{center}
\end{figure}
\begin{figure}[htbp]
\begin{center}
	\subfigure[]{\includegraphics[width=10.0cm]{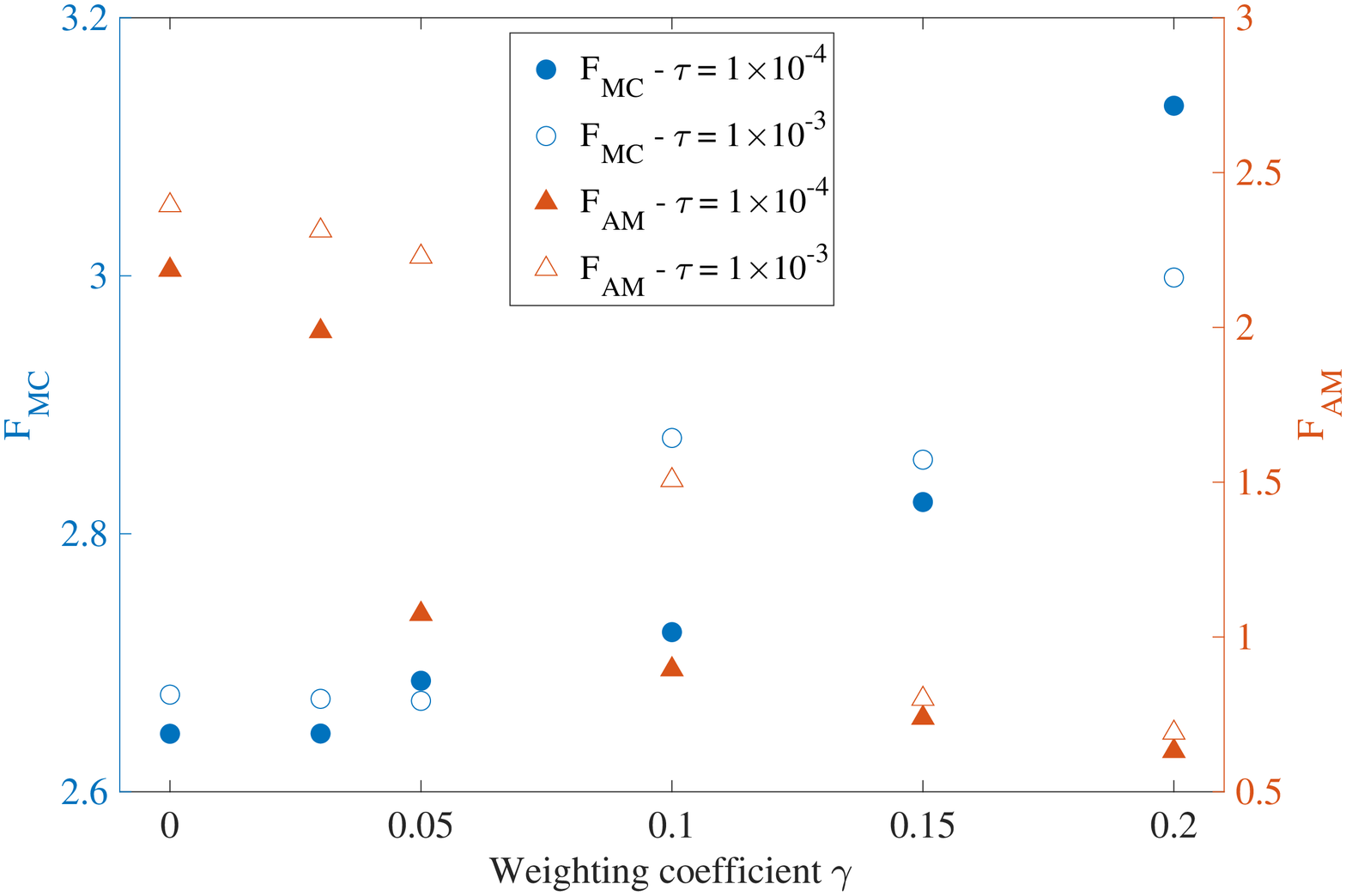}}
	%				\hspace{0.1cm}
	\subfigure[]{\includegraphics[width=10.0cm]{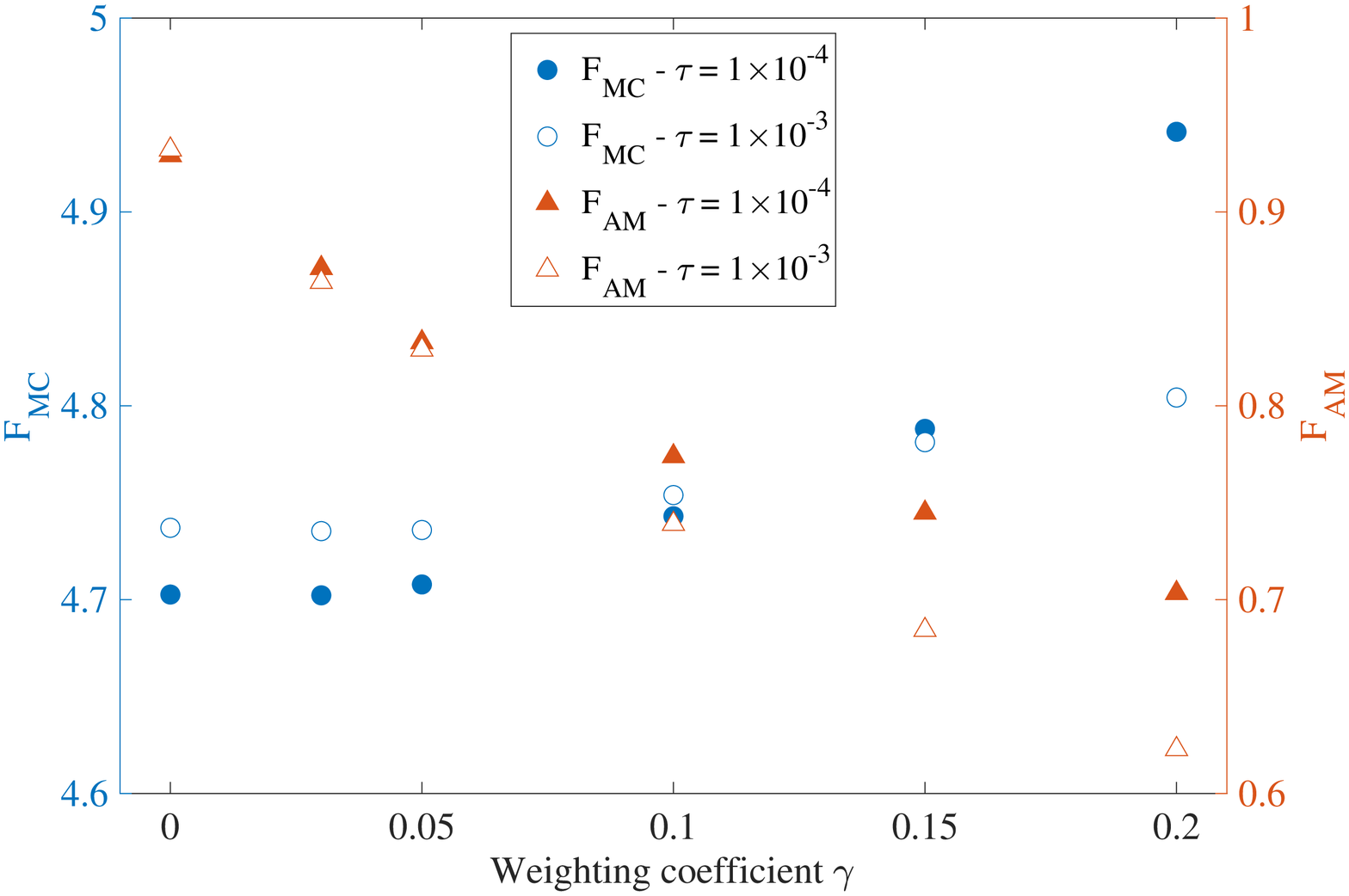}}
	\caption{Objective function with different weighting coefficient $\ga$: (a)cantilever model; (b)MBB beam model.}
	\label{fig:obj_ga}
\end{center}
\end{figure}
\begin{figure}[htbp]
\begin{center}
	\subfigure[]{\includegraphics[width=10.0cm]{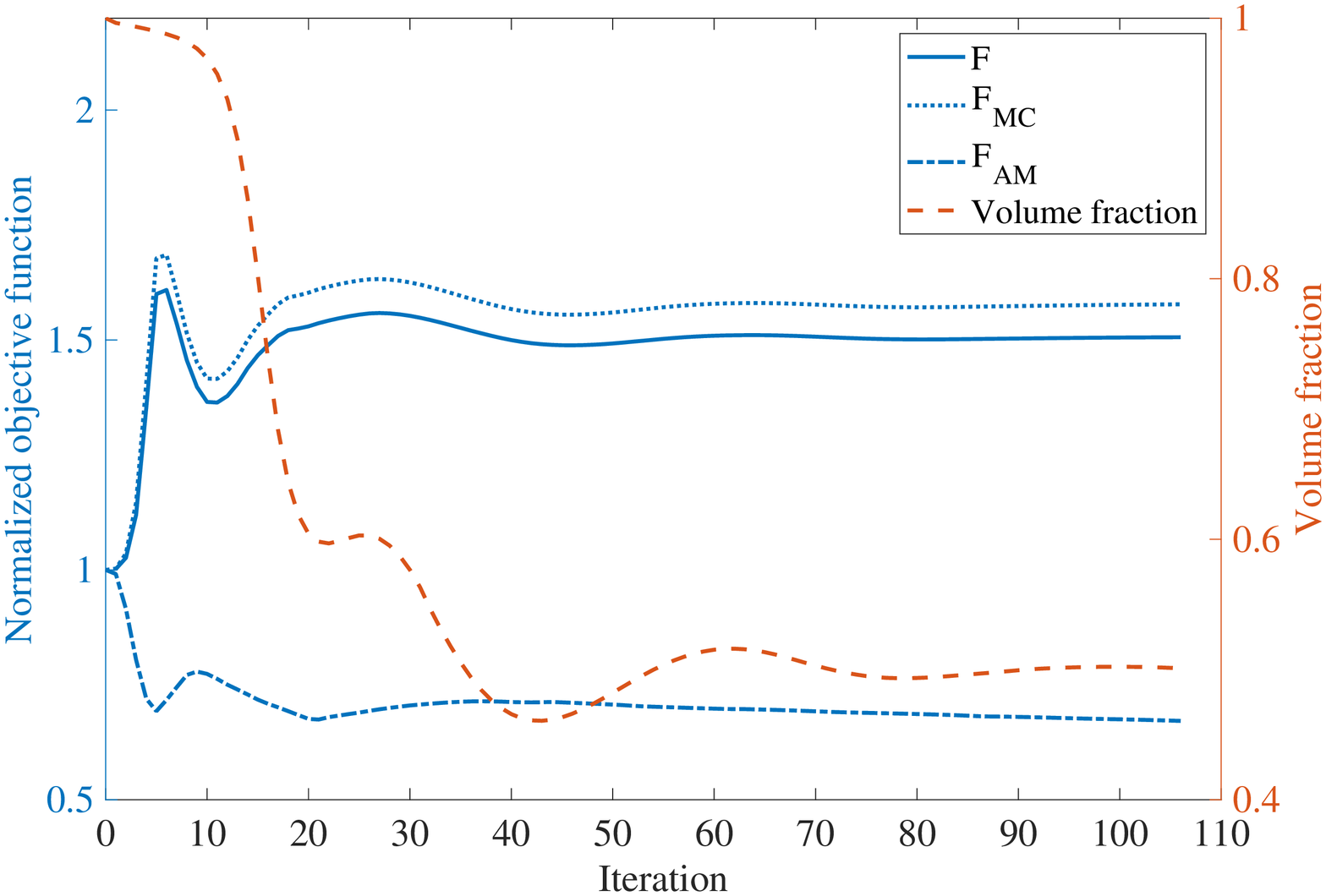}}
	%				\hspace{0.1cm}
	\subfigure[]{\includegraphics[width=10.0cm]{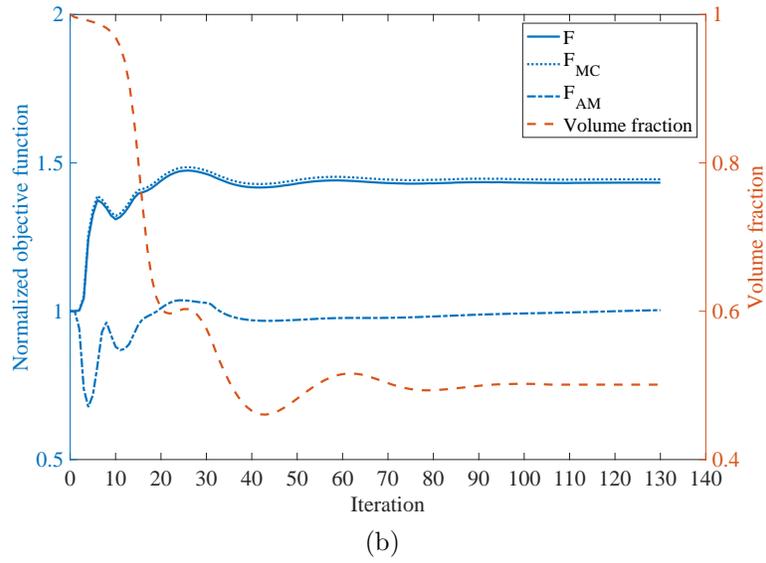}}
	\caption{Convergence history of the objective function and volume constraint for the case of $\ga = 0.1$: (a)cantilever model; (b)MBB beam model.}
	\label{fig:obj_vol}
\end{center}
\end{figure}
%\begin{figure}[htbp]
%	\begin{center}
%			\centering
%			\subfigure[]{\includegraphics[width=13.0cm]{Fig/AMdisp-Canti2.eps}}
%				\hspace{0.1cm}
%			\subfigure[]{\includegraphics[width=13.0cm]{Fig/AMdisp-MBB2.eps}}
%		\caption{Numerical results of part distortion induced by AM process: (a)cantilever model; (b)MBB beam model.}
%		\label{fig:AMdisp}
%	\end{center}
%\end{figure}
\begin{figure}[htbp]
\begin{center}
	\centering
	\subfigure[]{\includegraphics[width=13.0cm]{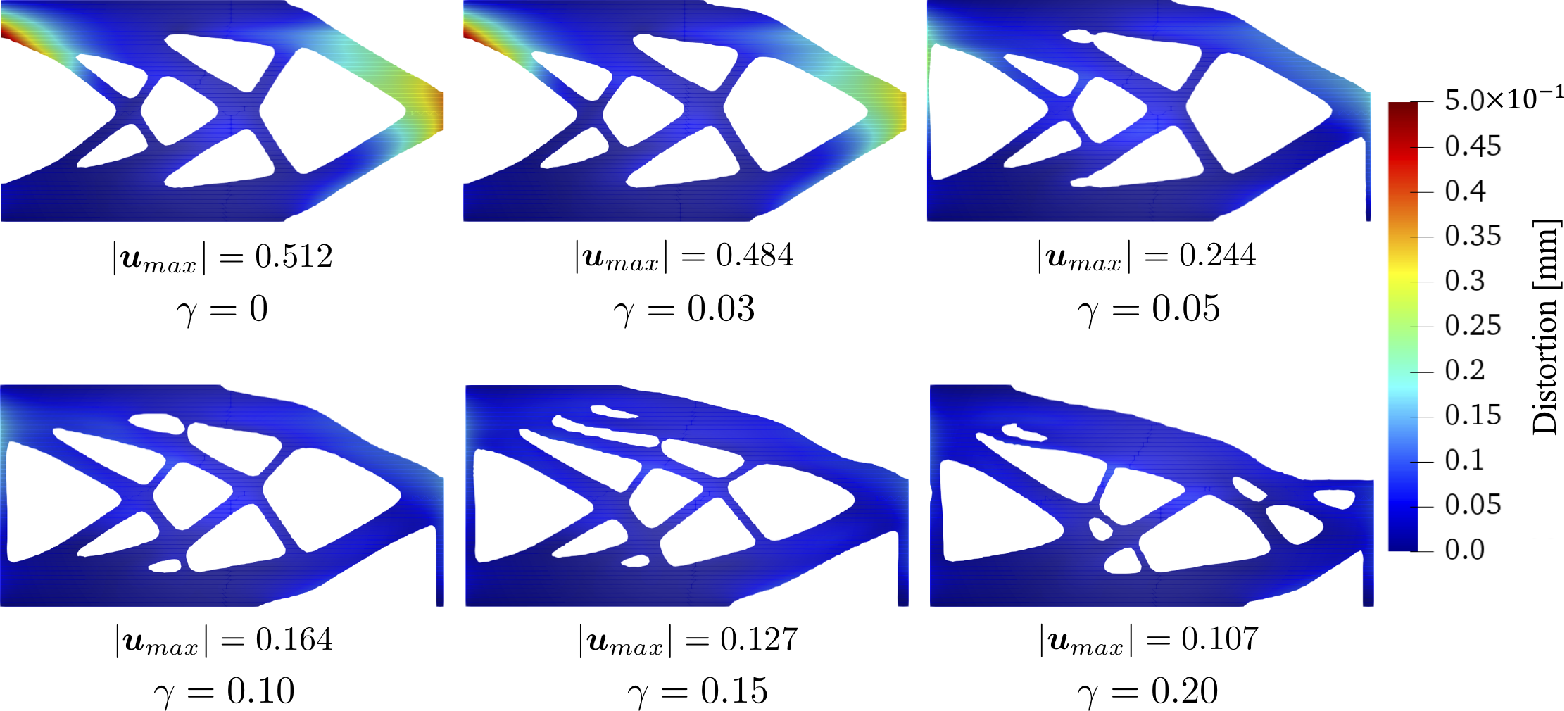}}
	%				\hspace{0.1cm}
	\subfigure[]{\includegraphics[width=13.0cm]{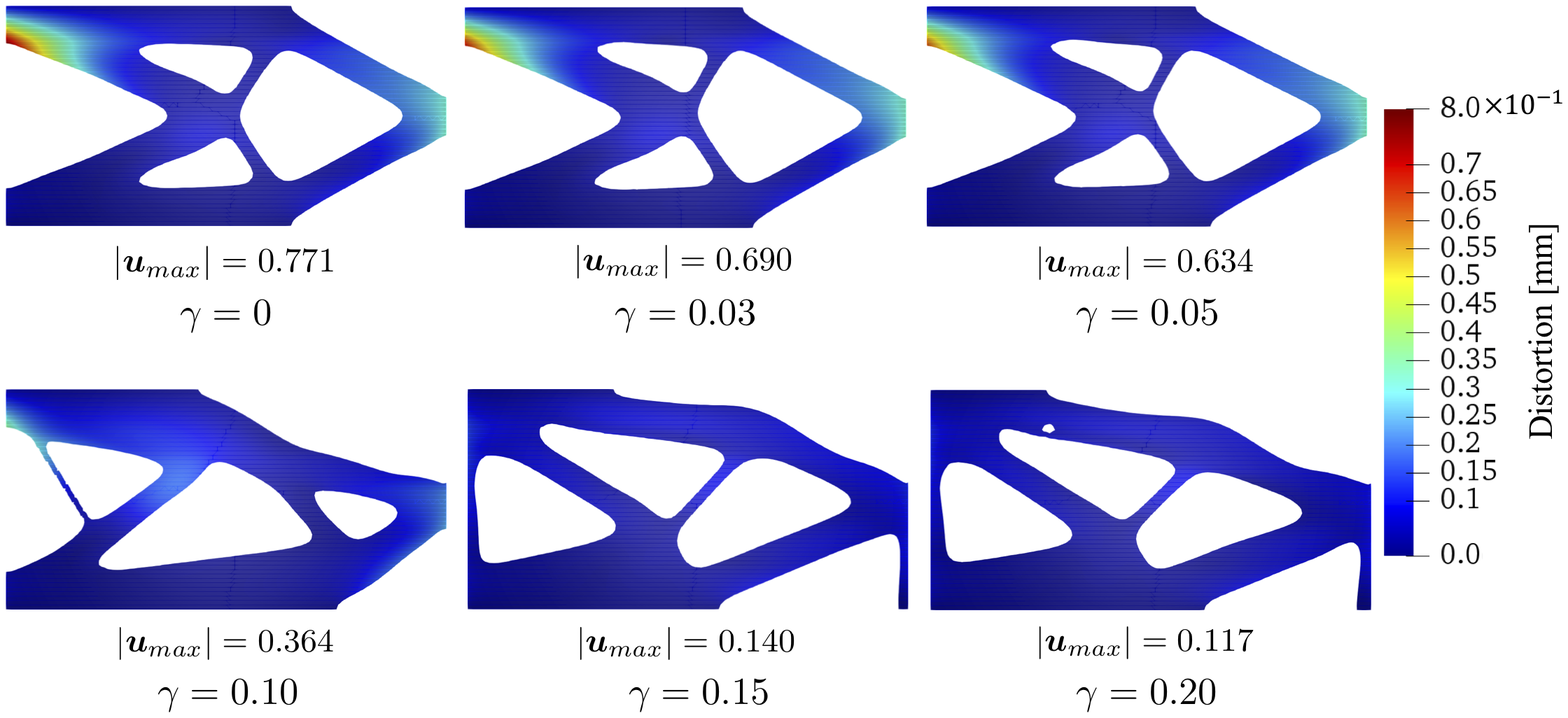}}
	\caption{Part distortion induced by the AM process for the cantilever model: (a)$\tau = 1\times10^{-4}$; (b)$\tau = 1\times10^{-3}$.}
	\label{fig:AMdispCanti} 
\end{center}
\end{figure}
\begin{figure}[htbp]
\begin{center}
	\centering
	\subfigure[]{\includegraphics[width=13.0cm]{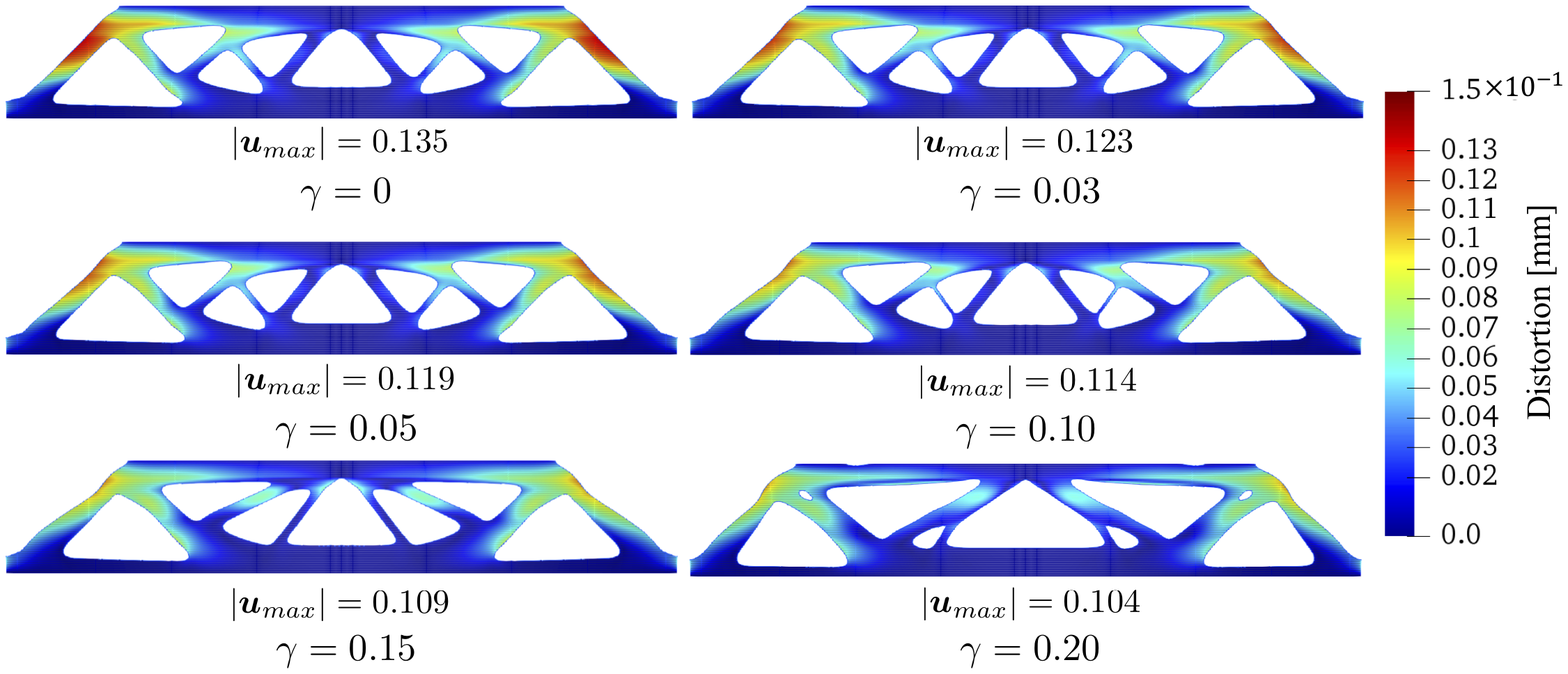}}
	%				\hspace{0.1cm}
	\subfigure[]{\includegraphics[width=13.0cm]{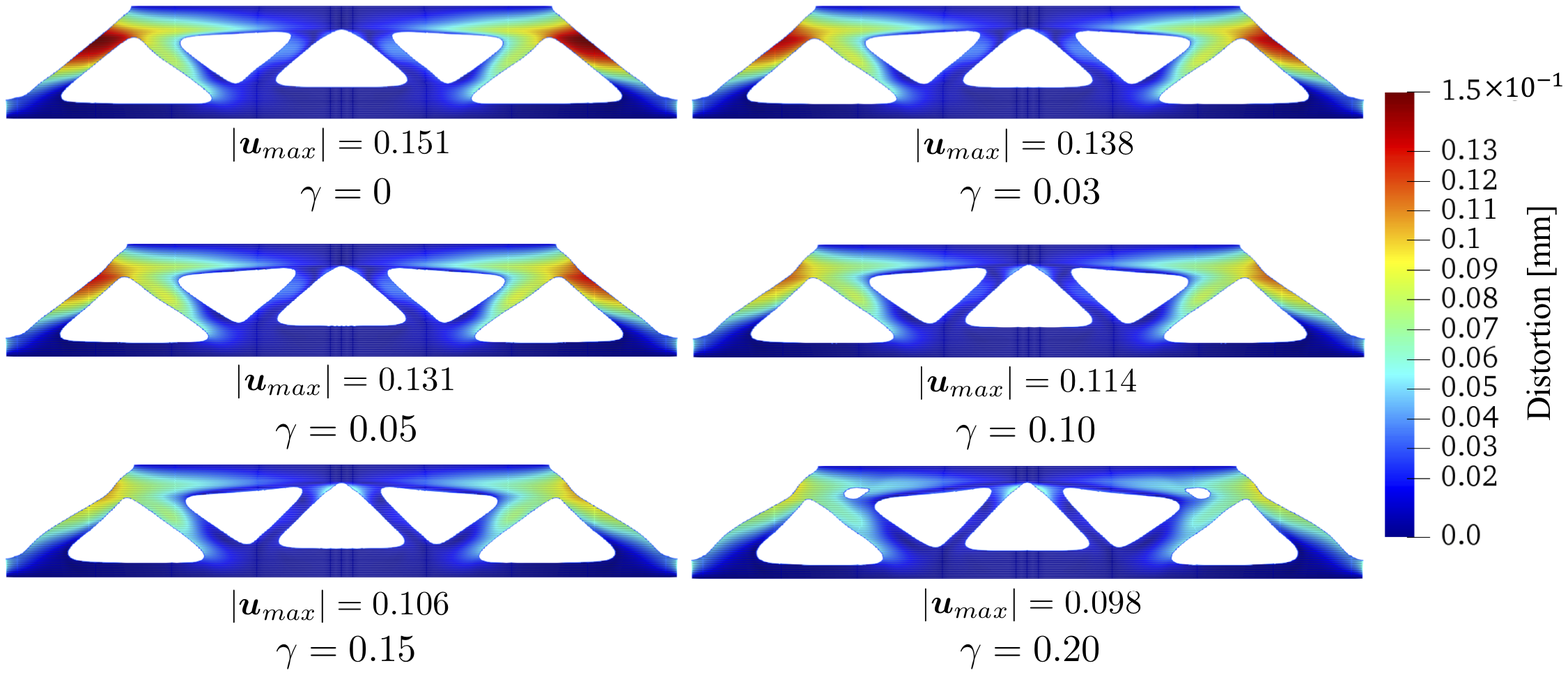}}
	\caption{Part distortion induced by the AM process for the MBB beam model: (a)$\tau = 1\times10^{-4}$; (b)$\tau = 1\times10^{-3}$.}
	\label{fig:AMdispMBB}
\end{center}
\end{figure}
Figures \ref{fig:resultga_Canti_t1e4}-\ref{fig:resultga_MBB_t1e3} show the evaluated optimal configurations for each model.
Figure \ref{fig:obj_ga} shows a plot of each objective function corresponding to the varying weighting coefficient.
Furthermore, $\ga = 0$ denotes that the objective function is only a compliance minimization problem.
Additionally, the geometrical complexity of the optimal configuration changes depending on the value of the regularization parameter $\tau$.
As shown in Fig. \ref{fig:obj_ga}, when the weighting coefficient $\ga$ is large, the objective function that represents the part distortion in AM decreases, and the compliance increases.
It can be seen that both numerical examples exhibit the same tendency, regardless of the regularization parameter, therefore, the designers can control the geometric complexity by adjusting $\tau$ first, followed by $\gamma$ to control the compliance and part distortion.
%Therefore, by adjusting $\ga$, compliance and part distortion can be controlled.
In other words, the part with the desired performance can be manufactured with high dimensional accuracy using AM.
If the target distortion $\bm{u}_{0}$ is replaced with $\bm{u} = \bm{u}-\bm{u}_{0}$ in Eq. \ref{eq:op1}, then the part can be manufactured with arbitrary dimensional accuracy.
Figure \ref{fig:obj_vol} shows the convergence history of the objective function and volume constraint for $\tau = 1\times10^{-4}$ and $\ga = 0.1$ for each model.
As the number of iterations increases, the objective function for the part distortion in AM decreases and compliance is maintained while satisfying the volume constraint.
Figures \ref{fig:AMdispCanti} and \ref{fig:AMdispMBB} show the numerical results of the part distortion induced by AM for the final shape.
%Figure \ref{fig:AMdisp} shows the numerical results of the part distortion induced by AM for the final shape.
The cantilever model has a protruding part outside the substrate where the distortion is large.
As $\ga$ increases, a structure appears in which the protrusions are fixed to the substrate to reduce the maximum displacement, and the displacement distribution becomes uniform.
In addition, since the MBB beam model has a structure fixed to the substrate, the distortion is smaller than that of the cantilever model.
Increasing $\ga$ changes the optimal configuration to reduce the maximum displacement.
%As $\ga$ increases, a support structure is created to reduce the maximum displacement, and the displacement distribution becomes uniform.
%As $\ga$ increases, the maximum displacement decreases, and the displacement distribution in the material domain becomes uniform.
This proves that the proposed methodology is efficient in considering the part distortion in AM.
\section{Conclusion}\label{sec:9}
In this study, we proposed a topology optimization method that considers the part distortion in AM, using a computationally inexpensive analytical model.
The main contributions of this study can be summarized as follows:
\begin{enumerate}
\item To predict the part-scale residual stress and distortion induced in the AM building process, the AM analytical model based on the inherent strain method and the identification method of the inherent strain component was proposed.
The experimentally identified in-plane inherent strain components and building process algorithm in the analytical model have been demonstrated to effectively predict the part-scale residual stress and distortion, without using coupled or nonlinear analysis.
The effect of the element size per layer in the analytical model on the accuracy and computational time was investigated, and the element size suitable for incorporation into topology optimization was proposed.
\item An objective function for reducing the part distortion in AM was proposed and a minimum mean compliance problem considering the part distortion was formulated. In the numerical implementation, an optimization algorithm was constructed and the non-dimensional sensitivity was used to enable simple adjustment of the weighting coefficient $\ga$.
\item In the minimum mean compliance problem, the proposed method provided an optimal configuration in which the compliance and part distortion in AM can be controlled by adjusting $\ga$ appropriately.
\end{enumerate}
In future work, we aim to construct a topology optimization method that can consider the entire AM process, including the overhang limitation and residual stress.
\section{Acknowledgments}\label{sec:10}
The authors are grateful to S. Nishiwaki and K. Furuta for useful discussions.
We would also like to thank K. Murata for technical assistance with X-ray stress measurement experiments.
This work was supported by the JSPS Grant for Scientific Research(B) JP19H02049.
\section*{References}
\bibliography{mybibfile}
\end{document}